\PassOptionsToPackage{numbers, sort&compress}{natbib}
\documentclass[iicol]{sn-jnl}

\usepackage{graphicx}%
\usepackage{multirow}%
\usepackage{amsmath,amssymb,amsfonts}%
\usepackage{amsthm}%
\usepackage{mathrsfs}%
\usepackage[title]{appendix}%
\usepackage{xcolor}%
\usepackage{textcomp}%
\usepackage{manyfoot}%
\usepackage{booktabs}%
\usepackage{algorithm}%
\usepackage{algorithmicx}%
\usepackage{algpseudocode}%
\usepackage{listings}%
\usepackage{amsmath, amssymb, booktabs}
\usepackage{subcaption}
\usepackage{hyperref}
\usepackage{natbib}
\usepackage{comment}
\usepackage{cuted}

\theoremstyle{thmstyleone}%
\theoremstyle{thmstyletwo}%

\theoremstyle{thmstylethree}%

\usepackage{scalerel}
\usepackage{tikz}
\usetikzlibrary{svg.path}

\definecolor{orcidlogocol}{HTML}{A6CE39}
\tikzset{
  orcidlogo/.pic={
    \fill[orcidlogocol] svg{M256,128c0,70.7-57.3,128-128,128C57.3,256,0,198.7,0,128C0,57.3,57.3,0,128,0C198.7,0,256,57.3,256,128z};
    \fill[white] svg{M86.3,186.2H70.9V79.1h15.4v48.4V186.2z}
                 svg{M108.9,79.1h41.6c39.6,0,57,28.3,57,53.6c0,27.5-21.5,53.6-56.8,53.6h-41.8V79.1z M124.3,172.4h24.5c34.9,0,42.9-26.5,42.9-39.7c0-21.5-13.7-39.7-43.7-39.7h-23.7V172.4z}
                 svg{M88.7,56.8c0,5.5-4.5,10.1-10.1,10.1c-5.6,0-10.1-4.6-10.1-10.1c0-5.6,4.5-10.1,10.1-10.1C84.2,46.7,88.7,51.3,88.7,56.8z};
  }
}

\newcommand\orcidicon[1]{\href{https://orcid.org/#1}{\mbox{\scalerel*{
\begin{tikzpicture}[yscale=-1,transform shape]
\pic{orcidlogo};
\end{tikzpicture}
}{|}}}}

\begin{document}

\title[Naked Singularities and Black Hole Thermodynamics from Shadows and Photon Rings in the Quantum Oppenheimer-Snyder Model]{\vspace{-1.6cm}Naked Singularities and Black Hole Thermodynamics from Shadows and Photon Rings in the Quantum Oppenheimer-Snyder Model}


\author[1]{\fnm{Jose Miguel} \sur{Ladino}\textsuperscript{\orcidicon{0000-0001-9812-4949}}}\email{miguel.ladino@correo.nucleares.unam.mx}

\author[1,2,3]{\fnm{Anuar} \sur{Idrissov}\textsuperscript{\orcidicon{0000-0001-6920-0463}}}\email{anuar.idrissov@gmail.com}

\author[1]{\fnm{Carlos E.} \sur{Romero-Figueroa}\textsuperscript{\orcidicon{0000-0001-5548-7766}}}\email{carlosed.romero@correo.nucleares.unam.mx}

\author*[1,3,4]{\fnm{Hernando} \sur{Quevedo}\textsuperscript{\orcidicon{0000-0003-4433-5550}}}\email{quevedo@nucleares.unam.mx}

\affil[1]{\orgname{Instituto de Ciencias Nucleares, Universidad Nacional Autónoma de México}, \orgaddress{\street{AP 70543}, \city{Mexico City}, \country{Mexico}}}

\affil[2]{\orgname{Fesenkov Astrophysical Institute}, \orgaddress{\street{Observatory 23}, \postcode{050020}, \city{Almaty}, \country{Kazakhstan}}}

\affil[3]{\orgname{Al-Farabi Kazakh National University}, \orgaddress{\street{Al-Farabi av. 71}, \postcode{050040}, \city{Almaty}, \country{Kazakhstan}}}

\affil[4]{\orgdiv{Dipartimento di Fisica and ICRA}, \orgname{Università di Roma ``La Sapienza''}, \orgaddress{\city{Roma}, \country{Italy}}}

\date{\today}

\abstract{We investigate the optical and thermodynamic properties of quantum Oppenheimer-Snyder (qOS) spacetimes in both the black hole and naked singularity regimes. By analyzing test-particle motion and employing numerical ray tracing of thin accretion disk models, we identify their characteristic optical signatures. We find that increasing the quantum correction parameter reduces the apparent shadow size. While qOS black holes exhibit dark inner shadows, naked singularities produce central bright rings, driven by an infinite repulsive core barrier, yet still retain key optical features such as photon rings. Furthermore, using a semi-analytical approach, we obtain higher-order photon-ring impact and gap parameters that closely align with our numerical findings. Notably, the gap parameter intervals for qOS naked singularities deviate from the Schwarzschild case, providing a concrete criterion for observational tests. 
Finally, in the black-hole sector, the application of  Geometrothermodynamics (GTD) identifies the specific-heat 
divergence at the Davies point.
 By linking photon-ring observables to GTD curvature scalars, we reveal a direct connection between qOS black hole shadows, thermodynamic stability, and microscopic interaction structures.}

\keywords{Black Hole Shadows,  Photon Rings, Black Hole Thermodynamics, Geometrothermodynamics}



\maketitle
\tableofcontents
\section{Introduction}\label{sec1}

Quantum corrections to black-hole spacetimes provide a useful framework for exploring possible deviations from classical general relativity in regimes where quantum-gravitational effects may become relevant. In particular, effective geometries inspired by Loop Quantum Gravity (LQG) have been proposed to capture quantum-geometry corrections while retaining a tractable black-hole description. Among these models, the quantum Oppenheimer--Snyder (qOS) spacetime provides a static and spherically symmetric geometry characterized by a quantum correction parameter $\hat{\alpha}$ \cite{qos0,qos1,qos2,qos3,qos4,qos5,qos6,qos7}. This solution has subsequently been investigated in several extensions, including the presence of a cosmological constant \cite{adsqos1,adsqos2,adsqos3} and rotation \cite{rotqos1,rotqos2,rotqos3,rotqos4}. More recently, the model has been extended to charged configurations~\cite{mazharimousavi2025charged,mazharimousavi2025phenomenological}, to higher dimensions with a cosmological constant~\cite{jiang2026higher}, to interiors described by loop quantum cosmology with a Lorentz term~\cite{ou2025lorentz}, and to asymmetric collapse~\cite{bobula2026asymmetric}, as well as to qOS black holes surrounded by quintessence, clouds of strings, or perfect fluid dark matter~\cite{ahmed2025quintessence,ahmed2026strings,sajjad2026qnm}. Its thermodynamic and radiative properties have been examined through particle tunneling, evaporation, canonical-ensemble stability, greybody factors, and heat engines in AdS backgrounds~\cite{tan2025tunneling,tan2025evaporation,chen2026canonical,lv2025greybody,javed2025heat}, while its phenomenology has been constrained with extreme mass-ratio inspirals, high-frequency quasi-periodic oscillations, and the S2 star orbit~\cite{qos1,rotqos1,errehymy2026joint}, and qOS primordial black holes have been proposed as dark matter candidates~\cite{wang2026primordial}. In addition, the tidal forces acting on radially infalling particles~\cite{idrissov2026tidal}, as well as thin accretion disks and gravitational capture cross sections of qOS black holes immersed in an external magnetic field~\cite{idrissov2026magnetic}, have recently been analyzed. Related LQG-inspired black-hole solutions with similar effective contributions have also been studied \cite{otherqos1,otherqos2,otherqos3,otherqos4,otherqos5,otherqos6,otherqos7}.

The optical properties of quantum-corrected geometries provide a complementary probe of their deviations from classical black holes. Null geodesics determine photon spheres, critical impact parameters, deflection angles, shadows, and photon rings, all linked to unstable photon trajectories \cite{Perlick_2022}. These optical observables can reveal modifications of the spacetime geometry and provide a direct link to potentially observable signatures of alternative gravity theories \cite{Fathi:2022pqv,Fathi:2024kce,Cruz:2004ts,ahmed2026constraining,fathi2024black,ahmed2026astrophysical,Ladino_2023,Antoniou2023,Vagnozzi_2023,Karthik2025,Zare_2026,Walia_2024,Lemos_2024,G_mez_2024,Tsukamoto_2024,Khodadi_2022}.

The observational relevance of black-hole shadows and photon rings has been highlighted by the Event Horizon Telescope (EHT), which has obtained horizon-scale images of the supermassive black holes M87$^\ast$ and Sgr A$^\ast$ \cite{EHTSgrA,EHTM87}. Photon rings arise from photons undergoing multiple passages around a compact object and are closely related to the structure of unstable photon orbits \cite{li2025images,gan2021photon}. Although current observations do not yet allow the photon-ring contribution to be separately resolved from the nearby lensed-ring structure \cite{li2025images}, their sensitivity to the strong-field spacetime geometry makes them promising probes of deviations from classical black-hole solutions \cite{cao2025image,rotqos2}. In this context, ray-tracing methods \cite{gelles2021role,cardenas2023adaptive} connect the accretion disk emission with the photon trajectories reaching the observer \cite{gralla2019black,li2025images}. While direct emission depends on the accretion flow, the photon-ring structure is governed primarily by the spacetime geometry and unstable photon orbits, providing a useful framework for testing different black-hole geometries \cite{Aratore2024, Tsupko_2022}. In fact, compact objects in alternative gravity theories and black hole mimickers, such as gravastars~\cite{Fauzi_2025,Rosa_2024}, wormholes~\cite{Chernov2023,Benavides2023}, and naked singularities~\cite{huang2024shadow,shaikh2019can,gyulchev2020observational,deliyski2025observing,Paul_2020,Tsukamoto_2021,Shaikh_2019,Chen2024,Mishra2026, Krajewski_2026,lake1991naked,ong2020space}, can exhibit distinctive ring-like structures. While their spacetime geometries are often nearly indistinguishable from those of standard black holes, the presence of a photon sphere can generate multiple photon rings, whereas configurations lacking one yield images composed entirely of direct emission \cite{huang2024shadow,shaikh2019can,gyulchev2020observational,deliyski2025observing}. For this reason, in addition to black holes, we will also explore the optical appearance of naked singularities. These objects have garnered significant interest, as they could offer an observational window into novel gravitational effects without the shielding of an event horizon, and their existence would challenge the cosmic censorship conjecture~\cite{huang2024shadow,shaikh2019can,gyulchev2020observational,deliyski2025observing,Paul_2020,Tsukamoto_2021,Shaikh_2019,Chen2024,Mishra2026, Krajewski_2026,lake1991naked,ong2020space}.

Furthermore, black hole thermodynamics has played a central role in establishing the connection between gravity, thermodynamics, and quantum theory. The identification of black hole entropy with the horizon area and the subsequent derivation of Hawking radiation established the thermodynamic character of black holes \cite{bekenstein1973black,hawking1975particle}.  Geometrothermodynamics (GTD) introduces a Legendre-invariant geometric framework for formulating equilibrium thermodynamics and describing thermodynamic interactions \cite{quevedo2007geometrothermodynamics}. In this approach, Legendre-invariant metrics defined on the thermodynamic phase space induce metrics on the equilibrium space, allowing resulting curvature singularities to be directly compared with established thermodynamic response functions and stability boundaries. This provides an alternative and rigorous method to investigate the thermodynamic and microscopic properties of black holes and their relationship to their observable optical signatures.
GTD has been successfully applied to a broad spectrum of physical systems, including ideal and real gases \cite{quevedo2022geometrothermodynamics,quevedo2011phase,quevedo2023geometrothermodynamic,romero2026}, ideal quantum gases \cite{zaldivar2023ideal}, magnetic materials \cite{QuevedoQuevedoSanchez2026}, the Ising model \cite{BravettiLopezMonsalvoNettelQuevedo2014}, chemical reactions \cite{quevedo2014geometric}, econophysical systems \cite{2023IJGMM..2050057Q,e28070762}, cosmological models \cite{romero2026quasi,e25060944,PhysRevD.86.063508,2019EPJC...79..577B,2014GrCo...20..208Q,Sanchezz2026}, and black holes across numerous alternative theories of gravity~\cite{quevedo2008geometrothermodynamics,Ladino:2024ned,ladino2025phase,LarranagaMojica2011,Gogoi,larranaga2012geometric,ref54,ref39,TharanathSureshKuriakose2015,Zhang2018GTD,ChannuieMomeni2018,GhaffariLuciano2025,Kurihara2018,Sanchez2016,QuevedoQuevedoSanchez2020,HuChenWang2017,GanChenWang2016,2025IJGMM..2250049J,SekhmaniEtAl2025,TajQuevedo2012,MoLiu2014,QuevedoSanchez2009,LucianoSheykhi2023,LucianoSaridakis2023,2023EPJC...83..710R,Beissen2024,JankeJohnstonKenna2010,quevedo2016geometrothermodynamics,HendiNaderi2015,HanChen2012,BravettiMomeniMyrzakulovAltaibayeva2013,Quevedo:2011fk,RizwanKumaraRajaniVaidAjith2019,quevedo2016einstein,ahmed2026shadow,Ladino:2026dru,quevedo2019quasi,Maya2026,Hernando2026,SHAH2026117662,Baghbani2026,QUEVEDO2026117321,Alessa2026,Biswas2026,Gogoi2026,BiswasSatyajit,QUEVEDO2026170601}.

Motivated by the interplay between spacetime geometry and black-hole thermodynamics, we investigate the optical and thermodynamic properties of qOS spacetimes in both the black hole and naked singularity regimes. Using numerical ray tracing of thin accretion disks \cite{Wang_2022, Yang_2023, Wang_2023, Chen2026, Zeng2023}, we analyze how the quantum correction parameter $\hat{\alpha}$ affects their characteristic optical signatures, including shadow size and photon-ring structures. We complement this with a semi-analytical approach, introduced in \cite{Aratore2024, Tsupko_2022}, to extract higher-order photon-ring gap parameters, linking these optical features to the thermodynamic stability and microscopic structure of the qOS spacetime within a quasi-homogeneous GTD framework \cite{quevedo2023unified,quevedo2019quasi}.

The paper is organized as follows. Section~\ref{sec2} introduces the qOS spacetime and its fundamental thermodynamic properties. Section~\ref{sec3} focuses on the optical signatures: we analyze test-particle motion, perform numerical ray-tracing simulations to contrast the visual appearance of qOS black holes against naked singularities, and compute higher-order impact and gap parameters in the strong-deflection limit. In Section~\ref{sec4}, we evaluate the GTD metrics and their corresponding Ricci scalars to characterize microscopic interactions, identifying specific heat divergences at the Davies point and establishing a direct connection between photon-ring observables and the thermodynamic behavior of the qOS spacetime. Section~\ref{sec5} summarizes our main conclusions.

\section{Quantum Oppenheimer-Snyder Model}\label{sec2}
We consider the qOS metric from
\cite{qos0}, which arises from incorporating
LQG corrections into the Oppenheimer--Snyder model of gravitational
collapse \cite{oppenheimer1939continued}. In this construction, the
collapsing dust interior is described by effective FLRW dynamics,
whose quantum-corrected evolution is matched to a static exterior
geometry \cite{piechocki2020quantum}. This leads to an effective
black-hole spacetime with quantum-geometric corrections to the
classical Schwarzschild geometry. The resulting static and spherically symmetric line element is given by
\begin{equation}
\begin{aligned}
ds^2 &= -f(r)\,dt^2 + f(r)^{-1}\,dr^2 + r^2d\Omega^2,\\
f(r) &= 1-\frac{2M}{r}+\frac{\hat{\alpha}M^2}{r^4}.
\end{aligned}
\label{eq:QOS_metric}
\end{equation}
Working in units $G=c=1$, the black hole mass $M$ has dimensions of length, whereas $\hat{\alpha}=16\sqrt{3}\pi\gamma^3\ell_P^2$ is the LQG correction parameter with dimensions of length squared, where $\ell_P$ is the Planck length and $\gamma$ is the dimensionless Barbero--Immirzi parameter \cite{rovelli1998immirzi}. Thus, $\hat{\alpha}$ encodes effective quantum-geometric effects and characterizes the deformation of Schwarzschild spacetime \cite{qos0}. Although not a conventional conserved charge, it acts as a hair-like quantum parameter, analogous to the concept of quantum hair \cite{coleman1992quantum}.

\subsection{Black Holes and Naked Singularities}

The possibility of naked singularities is an important aspect of gravitational theories \cite{lake1991naked}, as their existence is closely related to the cosmic censorship conjecture, which posits that spacetime singularities should remain hidden behind event horizons \cite{ong2020space}. In quantum-corrected black-hole models, modifications to the spacetime geometry can alter the horizon structure and potentially lead to configurations in which the singularity is not shielded by an event horizon \cite{fabbri2006semiclassical,casals2019quantum,malafarina2017classical}. To fully characterize the parameter space of the qOS solution, it is therefore essential to explore both black-hole and naked-singularity configurations. The outer event horizon radius, $r_h$, is determined by the largest real root of the equation
\begin{equation}
r_h^4 - 2M r_h^3 + \hat{\alpha} M^2 = 0.
\label{eq:horizon}
\end{equation}
For small deviations from the Schwarzschild geometry ($\hat{\alpha} \ll M^2$), a perturbative solution yields
\begin{equation}
r_h = 2M - \frac{\hat{\alpha}}{8M} - \frac{3\hat{\alpha}^2}{128M^3} - \frac{15\hat{\alpha}^3}{2048M^5} + \mathcal{O}(\hat{\alpha}^4).
\label{eq:rh_expansion}
\end{equation}
As $\hat{\alpha}$ increases, $r_h$ shrinks until $f(r)$ exhibits a degenerate root at the extremal state, yielding the critical quantities
\begin{equation}
\hat{\alpha}_{ext} = \frac{27}{16}M^2=1.6875M^{2}, \qquad r_h^{ext} = \frac{3}{2}M.
\label{eq:extremal_values}
\end{equation}
Thus, depending on the quantum parameter $\hat{\alpha}$, three distinct spacetime regimes are defined:
\begin{itemize}
\item \textbf{qOS black hole} ($0 < \hat{\alpha}/M^2 < 27/16$): The equation admits two positive roots, corresponding to an inner horizon and an outer event horizon.
\item \textbf{Extremal qOS black hole} ($\hat{\alpha}/M^2 = 27/16$): The inner horizon and outer event horizon merge into a single degenerate boundary.
\item \textbf{qOS naked singularity} ($\hat{\alpha}/M^2 > 27/16$): No real positive roots exist; consequently, no horizons form, leaving the central singularity exposed.\end{itemize}

In the classical limit $\hat{\alpha}=0$, the inner horizon
disappears and the Schwarzschild horizon is recovered. At short distances, the quantum correction dominates the metric
function, $f(r)\sim\hat{\alpha}M^2/r^4$. In fact, the Ricci scalar is
given exactly by
$R=-6\hat{\alpha}M^2/r^6$, and therefore diverges as $r\to0$ for
$\hat{\alpha}\neq0$. Moreover, the Kretschmann scalar behaves as
$K\sim468\hat{\alpha}^2M^4/r^{12}$, providing an additional
confirmation of the singularity. Tidal forces acting on radially infalling particles in the qOS black hole reverse sign in the interior, where the particles reach a turning point inside the Cauchy horizon~\cite{idrissov2026tidal}. Thus, the quantum
correction modifies the near-origin geometry without removing the
central singularity. Consequently, the qOS spacetime remains singular
at the center, in contrast to other black hole
solutions that can admit regular cores
\cite{ahmed2026shadow,RomeroFigueroa2026}. qOS naked singularities can exhibit either a photon sphere together with an anti-photon sphere, or no photon sphere at all, leading to distinct strong-field signatures. Unlike qOS black holes, reflective naked singularities lack a dark inner shadow: their infinite repulsive core prevents photon trapping and produces bright central rings instead. These features can distinguish them from black holes \cite{Paul_2020,Tsukamoto_2021,Shaikh_2019,Chen2024,Fauzi_2025,deliyski2025observing,Rosa_2024}.

\subsection{Black Hole Thermodynamics}\label{sec:thermo}

The qOS black hole is treated here as a thermodynamic system whose equilibrium
states are labeled by the entropy $S$ and by the quantum parameter
$\hat\alpha$. Following the quasi-homogeneous formulation of black hole
thermodynamics \cite{quevedo2019quasi, quevedo2017homogeneity,
quevedo2023unified}, $\hat\alpha$ is not a fixed coupling constant but a genuine
thermodynamic degree of freedom, so that the equilibrium space is
two-dimensional, $E^{a}=(S,\hat\alpha)$. Then, using the horizon condition $f(r_h)=0$ together with the Bekenstein--Hawking
entropy $S=\pi r_h^{2}$ \cite{bekenstein1973black, hawking1975particle}, the
black hole mass $M=M(S, \hat{\alpha})$ can be expressed as the fundamental thermodynamic potential
\begin{equation}
M= \frac{S^{3/2} - S \sqrt{S - \pi\hat{\alpha}}}{\hat{\alpha} \pi^{3/2}}.
\label{eq:Mass_Entropy}
\end{equation}
Throughout this section it is convenient to abbreviate
\begin{equation}
	\sigma \equiv \sqrt{S-\pi\hat\alpha}\,,
	\qquad
	S \geq \pi\hat\alpha \,,
	\label{eq:sigma}
\end{equation}
which renders all the expressions below compact.

The first law of thermodynamics in the extended space $(S,\hat\alpha)$ reads
$dM=T\,dS+A\,d\hat\alpha$, with 
\begin{equation}
	T=\left(\frac{\partial M}{\partial S}\right)_{\hat\alpha},
	\qquad
	A=\left(\frac{\partial M}{\partial \hat\alpha}\right)_{S},
	\label{eq:conjugates}
\end{equation}
where $T=T(S,\hat\alpha)$ is the temperature and $A=A(S,\hat\alpha)$ is the variable conjugate to the quantum parameter. Explicitly,
\begin{align}
	T = \frac{3\sqrt{S}\,\sigma-3S+2\pi\hat\alpha}
	{2\pi^{3/2}\hat\alpha\,\sigma},\qquad A = \frac{\sqrt{\pi}\,S}
	{2\,\sigma\left(\sqrt{S}+\sigma\right)^{2}} .
	\label{eq:Temperature}
\end{align}
We note that $T$ in Eq.~\eqref{eq:conjugates} is the temperature conjugate to the Bekenstein--Hawking entropy in the extended space $(S,\hat\alpha)$. Because the quantum correction in $f(r)$ scales as $M^2$, it does not coincide with the surface-gravity temperature $T_H=f'(r_h)/4\pi$; one finds $T_H=(\sigma/\sqrt{S})\,T$, so both vanish at extremality and agree in the Schwarzschild limit. Imposing the first law with $T_H$ instead requires a logarithmically corrected entropy and moves the Davies point to $r_h\simeq1.7599\sqrt{\hat\alpha}$~\cite{chen2026canonical}. Throughout this work, we adopt the Bekenstein--Hawking entropy and its conjugate temperature, which define the thermodynamic equilibrium space used in the subsequent stability and GTD analyzes. Consequently, the thermodynamic results presented below, including the location and geometric characterization of the Davies point, refer to this thermodynamic description.

The fundamental relation Eq.~\eqref{eq:Mass_Entropy} is not homogeneous, but it is
quasi-homogeneous \cite{quevedo2019quasi, quevedo2017homogeneity,
quevedo2023unified, Quevedoquasi}: under the rescaling
$S\to\lambda^{\beta_{S}}S$ and
$\hat\alpha\to\lambda^{\beta_{\hat\alpha}}\hat\alpha$, the radical
$\sqrt{S-\pi\hat\alpha}$ transforms homogeneously only if
$\beta_{\hat\alpha}=\beta_{S}$ and both terms of the numerator then scale as
$\lambda^{3\beta_{S}/2}$, while the prefactor $\hat\alpha^{-1}$ contributes
$\lambda^{-\beta_{S}}$, so that $M\to\lambda^{\beta_{S}/2}M$. Moreover, by choosing the scaling weights as follows, the fundamental equation $M(S,\hat\alpha)$ becomes quasi-homogeneous of
degree one, i.e.,
\begin{equation}
	\beta_{\hat\alpha}=\beta_{S},\quad
	\beta_{M}=\tfrac{1}{2}\beta_{S}
	\quad\Longrightarrow\quad\beta_{S}=\beta_{\hat\alpha}=2,\quad \beta_{M}=1 .
	\label{eq:betas}
\end{equation}
This differs from the Reissner--Nordstr\"om--AdS and Kerr--AdS cases, where
$\beta_{l}=\tfrac{1}{2}\beta_{S}$ \cite{Ladino:2024ned, ladino2025phase}: here
$\hat\alpha$ has dimensions of length squared, and indeed
$\beta_{\hat\alpha}=2\beta_{l}=\beta_{S}$. The quasi-homogeneous Euler identity
$\sum_{a}\beta_{a}E^{a}\partial_{a}M=\beta_{M}M$ \cite{quevedo2017homogeneity,
quevedo2023unified, mancilla2024generalized} then yields the Smarr formula
\cite{smarr1973mass} of the qOS black hole,
\begin{equation}
	M=2TS+2A\hat\alpha \,.
	\label{eq:smarr}
\end{equation}
Unlike asymptotically flat spacetimes, spacetimes with non-trivial asymptotics may yield different Smarr structures depending on the thermodynamic parameter space \cite{romero2024extended1}. Eq.~\eqref{eq:smarr} requires treating $\hat\alpha$ as a thermodynamic variable; freezing it incorrectly yields $M=2TS$.  Local thermal stability and phase structure are governed by response functions given by the second derivatives of the thermodynamic potential \cite{romero2024extended1, mansoori2015hessian, callen1998thermodynamics}. Using $M_{,xy} \equiv \frac{\partial^2 M}{\partial x \partial y}$, for the qOS black hole these read,
\begin{align}
	M_{,SS}&=\frac{\sqrt{S}\,\sigma-2S+3\pi\hat\alpha}
	{4\sqrt{\pi}\,\sqrt{S}\,\sigma^{3}\left(\sqrt{S}+\sigma\right)},\label{eq:MSS}\\[2pt]
	M_{,S\hat\alpha}&=-\frac{\sqrt{\pi}\left(2\sqrt{S}\,\sigma-S+2\pi\hat\alpha\right)}
	{4\,\sigma^{3}\left(\sqrt{S}+\sigma\right)^{2}},\label{eq:MSa}\\[2pt]
	M_{,\hat\alpha\hat\alpha}&=\frac{\pi^{3/2}S\left(\sqrt{S}+3\sigma\right)}
	{4\,\sigma^{3}\left(\sqrt{S}+\sigma\right)^{3}} .
	\label{eq:Maa}
\end{align}
    The relevant response functions are the heat capacity at fixed $\hat\alpha$, the heat capacity at fixed $A$, the isentropic compressibility $\kappa_S$, and the thermal expansion coefficient $\alpha_S$ \cite{romero2026geometric}:
\begin{align}
	C_{\hat\alpha}&=T\left(\frac{\partial S}{\partial T}\right)_{\hat\alpha}
	=\frac{T}{M_{,SS}},\label{eq:Cadef}\\[4pt]
	C_{A}&=T\left(\frac{\partial S}{\partial T}\right)_{A}=\frac{T\,M_{,\hat\alpha\hat\alpha}}
	{M_{,SS}M_{,\hat\alpha\hat\alpha}-M_{,S\hat\alpha}^{2}},\label{eq:CAdef}\\[4pt]
	\kappa_{S}&=\left(\frac{\partial\hat\alpha}{\partial A}\right)_{S}
	=\frac{1}{M_{,\hat\alpha\hat\alpha}},
	\qquad
	\alpha_{S}=\left(\frac{\partial\hat\alpha}{\partial T}\right)_{S}
	=\frac{1}{M_{,S\hat\alpha}} .
	\label{eq:response}
\end{align}
Inserting Eq.~\eqref{eq:Temperature} and Eqs.~\eqref{eq:MSS}--\eqref{eq:Maa} we obtain
their explicit form in the entropy representation,
\begin{align}
	C_{\hat\alpha}&=\frac{2\sqrt{S}\,\sigma^{2}\left(2\sigma-\sqrt{S}\right)}
	{\sqrt{S}\,\sigma-2S+3\pi\hat\alpha},\label{eq:Calpha}\\[2pt]
	\kappa_{S}&=\frac{4\,\sigma^{3}\left(\sqrt{S}+\sigma\right)^{3}}
	{\pi^{3/2}S\left(\sqrt{S}+3\sigma\right)},\label{eq:kappaS}\\[2pt]
	C_{A}&=\frac{2S\left(\sqrt{S}-2\sigma\right)\left(\sqrt{S}+3\sigma\right)}
	{\sigma\left(\sqrt{S}+4\sigma\right)} .\label{eq:CA}
\end{align}
The determinant of the Hessian of the potential, which controls $C_{A}$, is
\begin{equation}
	M_{,SS}M_{,\hat\alpha\hat\alpha}-\left(M_{,S\hat\alpha}\right)^{2}
	=-\frac{\pi\left(\sqrt{S}+4\sigma\right)}
	{16\,\sigma^{3}\left(\sqrt{S}+\sigma\right)^{4}} ,
	\label{eq:hessdet}
\end{equation}
and is therefore strictly negative on the whole physical domain. Likewise,
$M_{,\hat\alpha\hat\alpha}>0$ everywhere, so that $\kappa_{S}$ is finite and
non-vanishing and the entire phase structure is carried by $C_{\hat\alpha}$. Additionally, the temperature Eq.~\eqref{eq:Temperature} vanishes when $2\sigma=\sqrt{S}$, which
reproduces the degenerate-horizon condition $\hat{\alpha}_{ext}/M^{2}=27/16$ obtained
above and fixes the extremal configuration
\begin{equation}
	\begin{split}
		r_{h}^{\rm ext} &= \tfrac{2}{\sqrt{3}}\sqrt{\hat{\alpha}}
		\simeq 1.154701\sqrt{\hat{\alpha}},\\
		S_{\rm ext} &= \tfrac{4\pi\hat{\alpha}}{3}\simeq 4.188790\,\hat\alpha,\\
		M_{\rm ext} &= \tfrac{4}{3\sqrt{3}}\sqrt{\hat{\alpha}}
		\simeq0.769800\sqrt{\hat\alpha},
	\end{split}
	\label{eq:extremal}
\end{equation}
at which $T=C_{\hat\alpha}=C_{A}=0$, while
$\kappa_{S}=\tfrac{3\sqrt{3}}{5}\hat\alpha^{3/2}\simeq1.039230\,\hat\alpha^{3/2}$
remains finite. The Davies point \cite{davies1978thermodynamics, davies1977thermodynamic} is the
state at which $T$ attains its maximum value and, equivalently, at which
$C_{\hat\alpha}$ diverges,
\begin{equation}
	\left( \frac{\partial T}{\partial S} \right)_{\hat{\alpha}} = 0, \quad
	\left( \frac{\partial^2 T}{\partial S^2} \right)_{\hat{\alpha}} < 0
	\quad\Longleftrightarrow\quad M_{,SS}=0 .
	\label{eq:daviescond}
\end{equation}
From Eq.~\eqref{eq:MSS}, the condition $M_{,SS}=0$ is equivalent to the quadratic
equation $3S^{2}-11\pi\hat\alpha S+9\pi^{2}\hat\alpha^{2}=0$, whose only root
inside the physical domain is
\begin{equation}
	\begin{split}
		S_{m}&=\frac{11+\sqrt{13}}{6}\,\pi\hat{\alpha}\simeq7.647449\,\hat\alpha,
		\\
		r_{hm}&=\sqrt{\frac{\left(11+\sqrt{13}\right)\hat\alpha}{6}}\simeq1.560211\sqrt{\hat\alpha},
		\\
		M_{m}&\simeq0.882676\sqrt{\hat\alpha},
		\qquad
		\hat\alpha_{m}\simeq1.283504\,M^{2},
	\end{split}
	\label{eq:davies}
\end{equation}
with maximum temperature
\begin{equation}
	T_{m}=\frac{3\left(\sqrt{13}-2\right)}{2\pi\left(5+2\sqrt{13}\right)}
	\sqrt{\frac{6}{\left(11+\sqrt{13}\right)\hat\alpha}}
	\simeq\frac{0.040237}{\sqrt{\hat\alpha}},
	\label{eq:Tmax}
\end{equation}
$\kappa_{Sm}\simeq11.488626\,\hat\alpha^{3/2}$ and
$C_{Am}\simeq-8.652958\,\hat\alpha$. Here, and in what follows, the subscript
$m$ denotes the Davies point. Consequently, in the entropy representation the qOS black hole is defined on
\begin{equation}
	S \in [S_{\rm ext}, \infty)=[4.188790\,\hat{\alpha}, \infty),
	\quad
	S_m \simeq 7.647449\,\hat{\alpha},
	\label{eq:Srange}
\end{equation}
and the sign of the heat capacity splits this interval into two branches,
\begin{equation}
	C_{\hat\alpha}>0 \;\, \text{for} \;\, S_{\rm ext}\le S<S_{m},
	\quad
	C_{\hat\alpha}<0 \;\, \text{for} \;\, S>S_{m},
	\label{eq:branches}
\end{equation}
with $C_{\hat\alpha}\to-2S$ in the Schwarzschild limit $\hat\alpha\to0$. The
temperature therefore rises from zero at extremality to the single maximum
Eq.~\eqref{eq:Tmax} and decays back to zero as $S\to\infty$, so that
Eq.~\eqref{eq:branches} describes a locally stable small black hole branch
and a locally unstable large black hole branch, separated by a spinodal
point at $S_{m}$. This is precisely the Davies structure of the Reissner--Nordstr\"om
black hole and of the quantum-corrected cosmological horizon
\cite{romero2026quasi}. However, since the stationary point of $T$ is
a maximum here, the branch assignment is opposite to that in the
Schwarzschild--AdS and Kerr--AdS cases, where the stationary point is
a minimum \cite{Ladino:2024ned, ladino2025phase}. In Sec.~\ref{sec4}, we explore the microscopic interpretation of these two branches, their classification into Microscopic Thermodynamic Phases (MTPs), and their connection to shadow and photon ring observables.

\section{Shadow Images and Photon Rings}\label{sec3}

To analyze the propagation of test particles in the qOS spacetime and perform numerical backward ray-tracing, we employ the Hamiltonian formalism. For a spacetime characterized by the metric function $f(r)$,  the geodesic motion is given by
\begin{equation}
\mathcal{H} = \frac{1}{2} g^{\mu\nu} p_\mu p_\nu = \frac{1}{2} \left[ -\frac{p_t^2}{f(r)} + f(r) p_r^2 + \frac{p_\theta^2}{r^2} + \frac{p_\phi^2}{r^2 \sin^2\theta} \right],
\end{equation}
where $p_\mu$ is the canonically conjugate momentum. The motion is constrained by $\mathcal{H} = -\frac{1}{2} \delta$, where $\delta = 1$ for massive particles (timelike geodesics) and $\delta = 0$ for massless particles, such as photons (null geodesics).

Due to the spacetime symmetries, the geometry admits two Killing vector fields, $\partial_t$ and $\partial_\phi$, which guarantee the conservation of the generalized momenta corresponding to the test particle's energy, $p_t \equiv -E$, and axial angular momentum, $p_\phi \equiv L$. Consequently, these quantities remain constant along the trajectory, meaning $\dot{p}_t = \dot{p}_\phi = 0$, where the overdot denotes differentiation with respect to the affine parameter $\lambda$.

Using Hamilton's equations of motion, $\dot{x}^\mu = \frac{\partial \mathcal{H}}{\partial p_\mu}$ and $\dot{p}_\mu = -\frac{\partial \mathcal{H}}{\partial x^\mu}$, we derive the equation of motion for the spacetime coordinates as
\begin{equation}
\dot{t} = \frac{E}{f(r)}, \quad \dot{r} = f(r) p_r, \quad \dot{\theta} = \frac{p_\theta}{r^2}, \quad \dot{\phi} = \frac{L}{r^2 \sin^2\theta}.
\label{eqmotion1}
\end{equation}

Similarly, the dynamical equations of motion governing the evolution of the remaining momentum components are
\begin{align} 
\dot{p}_r &= -\frac{1}{2} \frac{f'(r)}{f(r)^2} E^2 - \frac{1}{2} f'(r) p_r^2 + \frac{p_\theta^2}{r^3} + \frac{L^2}{r^3 \sin^2\theta}, \label{eqmotion2} \\ 
\dot{p}_\theta &= \frac{\cos\theta}{\sin^3\theta} \frac{L^2}{r^2}. 
\label{eqmotion3}
\end{align}

Together, these expressions constitute a complete set of 8 coupled first-order ordinary differential equations. We solve this system to perform the numerical simulation of the black hole shadow images. 

\subsection{Motion of Test Particles}\label{sec3.1}


\begin{figure*}[h]
 \vspace{0.2cm}
\begin{minipage}[t]{0.475\linewidth}
 \centering
 \hspace{-7.2cm} \includegraphics[width=0.99\linewidth]{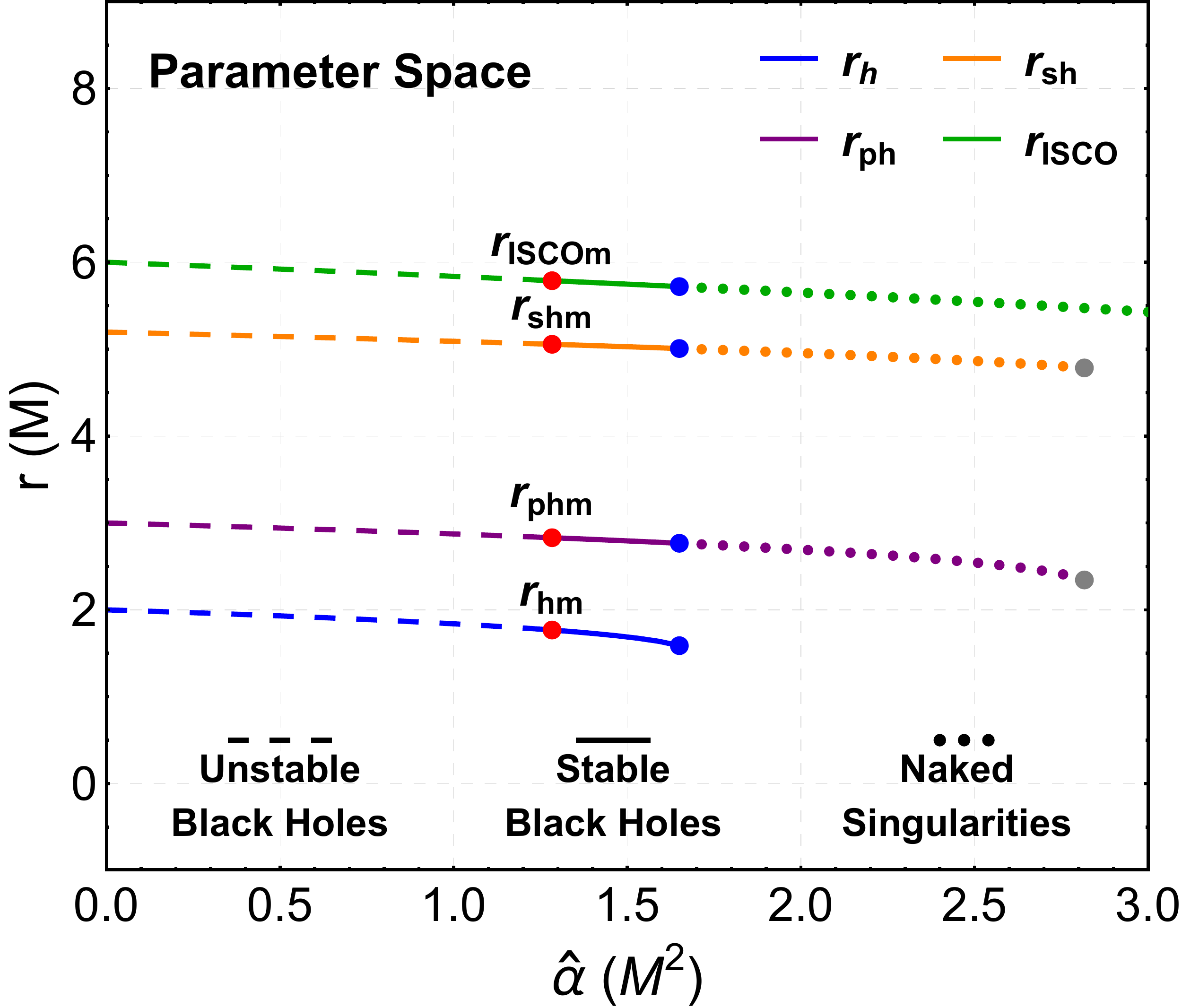}
  \hspace{-8cm} (a)
\end{minipage}%
\hfill%
\begin{minipage}[t]{0.49\linewidth}
 \centering
 \hspace{-7.7cm} \includegraphics[width=0.975\linewidth]{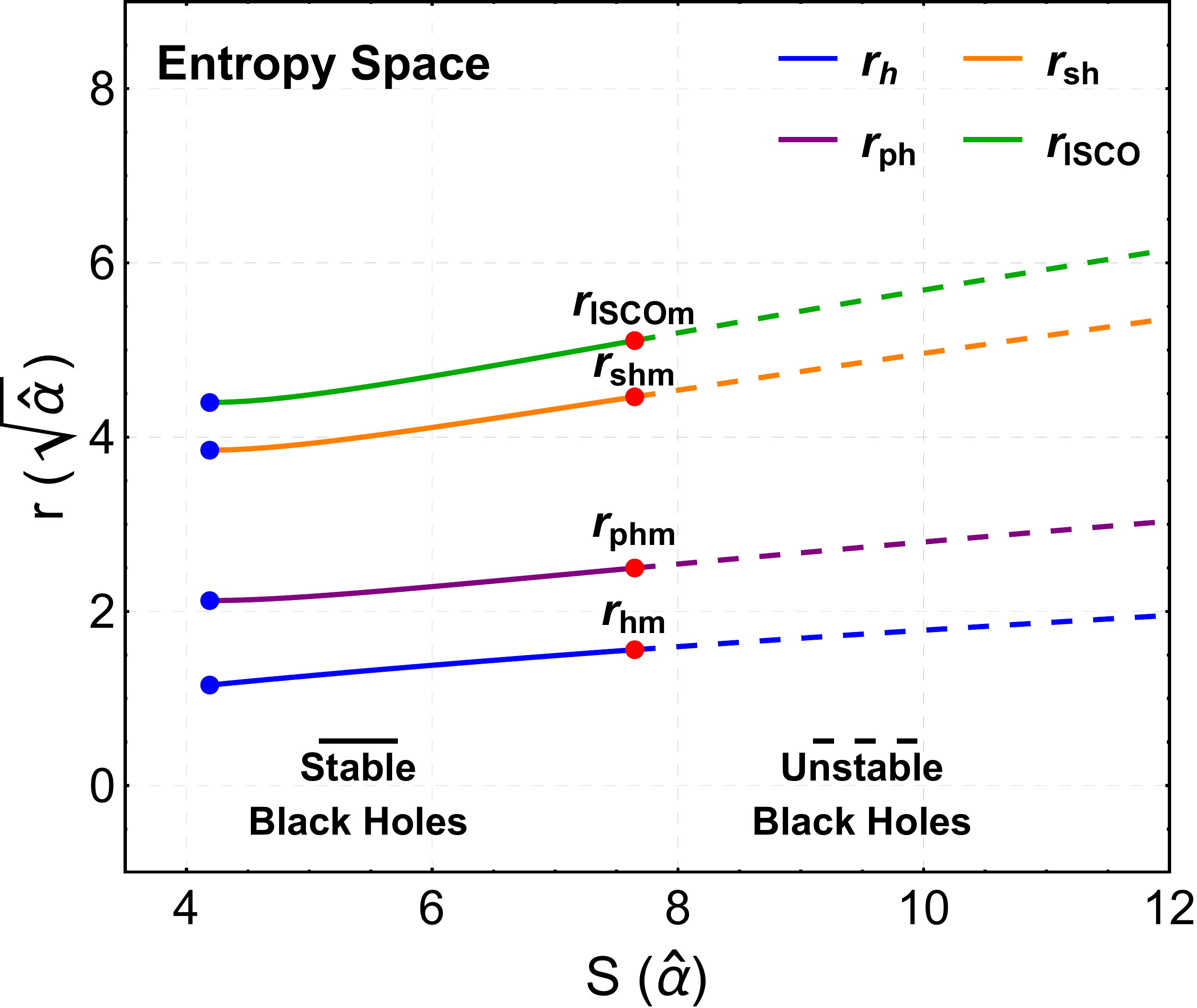}
 \hspace{-8cm} (b)
\end{minipage}%
\hfill \\
\vspace{-0.1cm}
\caption{Radii $r_h$, $r_{ph}$, $r_{sh}$, and $r_{ISCO}$ in (a) parameter and (b) entropy spaces for the quantum Oppenheimer-Snyder spacetime. Stable and unstable black hole and naked singularity regions are shown. Red dots mark the Davies point, with $r_{hm}$, $r_{phm}$, $r_{shm}$, and $r_{ISCOm}$ denoting the corresponding radii; blue dots indicate the extremal case. Gray dots in (a) mark where the photon sphere and critical shadow cease to exist for naked singularities.}

\label{ShadowProfiles1}
\end{figure*}

    
\begin{table*}[h]
\centering
\caption{Range and values at the Davies point for the event horizon ($r_h$), innermost stable circular orbit ($r_{ISCO}$), photon sphere radius ($r_{ph}$) and critical shadow radius ($r_{sh}$), calculated within the parameter and entropy spaces.}
\label{tab:unified-observables}
\setlength{\tabcolsep}{1.8pt} 
\renewcommand{\arraystretch}{1.0} 
\begin{tabular*}{\textwidth}{@{\extracolsep{\fill}}cc|cc}
\hline
\multicolumn{2}{c|}{\textbf{Range $[T=0, T= T_{Schw}]$}} & \multicolumn{2}{c}{\textbf{Value at Davies Point $(T=T_m)$}} \\ \hline\hline

\multicolumn{4}{c}{\textbf{Reissner-Nordström Black Holes}} \\ \hline
\textbf{Parameter Space} & \textbf{Entropy Space} & \textbf{Parameter Space} & \textbf{Entropy Space} \\
\textit{$r(M)$ vs $Q(M)$} & \textit{$r(Q)$ vs $S(Q^2)$} & \textit{$r(M)$ vs $Q(M)$} & \textit{$r(Q)$ vs $S(Q^2)$} \\ 
$Q \in [0, M]$ & $S \in [3.1416 Q^2, \infty)$ & $Q_m \approx 0.8660 M$ & $S_m \approx 9.4248 Q^2$ \\ \hline

$r_{h} \in [M, 2 M]$ & $r_{h} \in [Q, \infty)$ 
& $r_{hm} = 1.5 M$ & $r_{hm} \approx 1.7321 Q$ \\ \hline

$r_{ISCO} \in [4 M, 6 M]$ & $r_{ISCO} \in [4 Q, \infty)$ 
& $r_{ISCOm} \approx 4.6533 M$ & $r_{ISCOm} \approx 5.3732 Q$ \\ \hline

$r_{ph} \in [2 M, 3 M]$ & $r_{ph} \in [2 Q, \infty)$ 
& $r_{phm} \approx 2.3660 M$ & $r_{phm} \approx 2.7321 Q$ \\ \hline

$r_{sh} \in [4 M, 5.1962 M]$ & $r_{sh} \in [4 Q, \infty)$ 
& $r_{shm} \approx 4.4037 M$ & $r_{shm} \approx 5.0849 Q$ \\ \hline\hline

\multicolumn{4}{c}{\textbf{Quantum Oppenheimer-Snyder Black Holes}} \\ \hline
\textbf{Parameter Space} & \textbf{Entropy Space} & \textbf{Parameter Space} & \textbf{Entropy Space} \\
\textit{$r(M)$ vs $\hat{\alpha}(M^2)$} & \textit{$r(\sqrt{\hat{\alpha}})$ vs $S(\hat{\alpha})$} & \textit{$r(M)$ vs $\hat{\alpha}(M^2)$} & \textit{$r(\sqrt{\hat{\alpha}})$ vs $S(\hat{\alpha})$} \\ 
$\hat{\alpha} \in [0, 1.6875 M^2]$ & $S \in [4.1888 \hat{\alpha}, \infty)$ & $\hat{\alpha}_m \approx 1.2835 M^2$ & $S_m \approx 7.6474 \hat{\alpha}$ \\ \hline

$r_{h} \in [1.5 M, 2 M]$ & $r_{h} \in [1.1547 \sqrt{\hat{\alpha}}, \infty)$ 
& $r_{hm} \approx 1.7676 M$ & $r_{hm} \approx 1.5602 \sqrt{\hat{\alpha}}$ \\ \hline

$r_{ISCO} \in [5.7127 M, 6 M]$ & $r_{ISCO} \in [4.3976 \sqrt{\hat{\alpha}}, \infty)$ 
& $r_{ISCOm} \approx 5.7877 M$ & $r_{ISCOm} \approx 5.1087 \sqrt{\hat{\alpha}}$ \\ \hline

$r_{ph} \in [2.7589 M, 3 M]$ & $r_{ph} \in [2.1238 \sqrt{\hat{\alpha}}, \infty)$ 
& $r_{phm} \approx 2.8301 M$ & $r_{phm} \approx 2.4981 \sqrt{\hat{\alpha}}$ \\ \hline

$r_{sh} \in [5.0021 M, 5.1962 M]$ & $ r_{sh} \in [3.8506 \sqrt{\hat{\alpha}}, \infty)$ 
& $r_{shm} \approx 5.0560 M$ & $r_{shm} \approx 4.4628 \sqrt{\hat{\alpha}}$ \\ \hline

\end{tabular*}
\end{table*}

Without loss of generality, restricting the motion to the equatorial plane ($\theta = \pi/2$, $p_\theta = 0$) reduces the dynamics to the radial equation
\begin{equation}
\dot{r}^2 = E^2 - V_{\mathrm{eff}}(r), \qquad V_{\mathrm{eff}}(r) = f(r) \left( \delta + \frac{L^2}{r^2} \right).
\label{eq:rdot}
\end{equation}

For massive particles ($\delta=1$), circular orbits require $\dot{r}=0$ and $V_{eff}'(r)=0$, which determine the conserved quantities $L$ and $E$. The innermost stable circular orbit (ISCO) radius, $r_{ISCO}$, satisfying $V_{eff}''(r_{ISCO}) = 0$, can be found for qOS black holes via the condition:
\begin{align}
6 M r_{ISCO}^6 - r_{ISCO}^7 - 9 M^2 r_{ISCO}^3 \hat{\alpha} & \nonumber \\
- 4 M r_{ISCO}^4 \hat{\alpha} + 12 M^3 \hat{\alpha}^2 &= 0,
\label{eq:ISCO_condition}
\end{align}
whose root structure depends strongly on $\hat{\alpha}$. Unlike Kerr or Schwarzschild metrics where test particles can only rest at infinity, the qOS naked singularity features a zero-gravity sphere at $r_{zg} = (2M\hat{\alpha})^{1/3}$ where radial acceleration vanishes ($f'(r_{zg}) = 0$) \cite{Mishra2026, Krajewski_2026}. Inside $r_{zg}$, gravity becomes repulsive. This repulsive core alters orbital dynamics, creating three physical regimes:

\begin{itemize}
\item \textbf{qOS black hole} ($\hat{\alpha} \le \frac{27}{16} M^2$): Eq.~\eqref{eq:ISCO_condition} yields a unique positive root outside the horizon (the standard ISCO). Gravity is attractive outside the horizon and all inner circular orbits are unstable.

\item \textbf{qOS naked singularity with multiple stable orbits} ($\frac{27}{16} M^2 < \hat{\alpha} \le 5.35323 M^2$): The repulsive region yields up to three positive roots, creating two disconnected stable zones \cite{Mishra2026, Krajewski_2026}. The outermost root is the Ordinary Marginally Stable Circular Orbit (OMSCO), analogous to a standard ISCO, and the innermost root falls below $r_{zg}$ and is physically irrelevant. For $\frac{729}{256} M^2 < \hat{\alpha} \le 5.35323 M^2$, the intermediate root is the Extraordinary Marginally Stable Circular Orbit (EMSCO): orbits between the EMSCO and the OMSCO are unstable, while those outside the OMSCO and between $r_{zg}$ and the EMSCO are stable. For $\frac{27}{16} M^2 < \hat{\alpha} \le \frac{729}{256} M^2$, instead, the intermediate root lies between the anti-photon sphere $r_{aph}$ and the photon sphere $r_{ph}$ discussed below, where no timelike circular orbits exist; the inner stable zone then extends over $r_{zg}<r<r_{aph}$, and orbits with $r_{ph}<r<r_{\rm OMSCO}$ are unstable.

\item \textbf{qOS naked singularity with continuous stability} ($\hat{\alpha} > 5.35323 M^2$): The unstable band vanishes. The only root falls below $r_{zg}$ and is irrelevant, and consequently, massive-particle circular orbits are stable everywhere in the physical region.
\end{itemize}
For subsequent shadow imaging simulations of naked singularities, one of the emission models we consider is defined to start at the ISCO; in this case, we assume the accretion disk begins at the OMSCO (hereafter referred to as the ISCO).

For massless particles ($\delta=0$), introducing the impact parameter $b \equiv |L|/E$ and rescaling the affine parameter reduces Eq.~\eqref{eq:rdot} to
\begin{equation}
\left(\frac{dr}{d\tilde{\lambda}}\right)^2 = \frac{1}{b^2} - \tilde{V}_{\mathrm{eff}}(r), \qquad \tilde{V}_{\mathrm{eff}}(r) = \frac{f(r)}{r^2}.
\label{eq:rdot_photons}
\end{equation}
The photon sphere of qOS black holes consists of unstable circular photon orbits at $r = r_{ph}$, determined by $\tilde{V}_{\mathrm{eff}}'(r_{ph}) = 0$, yielding $r_{ph} f'(r_{ph}) - 2f(r_{ph}) = 0$, and thus
\begin{equation}
6 M r_{ph}^3 - 2 r_{ph}^4 - 6 M^2 \hat{\alpha} = 0.
\label{eq:photon_sphere}
\end{equation}
Then, $r_{ph}=r_{ph}(M,\hat{\alpha})$ is given by
\begin{align}
    r_{ph} &= \frac{3M}{4} + \frac{\sqrt{\mathcal{P}}}{2} + \frac{1}{4} \sqrt{27M^2 - 4\mathcal{P} + \frac{27M^3}{\sqrt{\mathcal{P}}}} \,,  \notag\\
    \mathcal{P} &= \frac{9M^2}{4} + \frac{4 \cdot 2^{1/3} M \hat{\alpha}^{2/3}}{\left( 27M + \sqrt{729M^2 - 256\hat{\alpha}} \right)^{1/3}} \label{eq:rph} \\& \quad + \frac{M}{2^{1/3}} \left[ \hat{\alpha} \left( 27M + \sqrt{729M^2 - 256\hat{\alpha}} \right) \right]^{1/3} \,. \notag
\end{align}
For an asymptotic observer ($r_0 \to \infty$), the critical impact parameter $b_c$, which satisfies $b_c = 1/\sqrt{\tilde{V}_{\mathrm{eff}}(r_{ph})}$, defines the critical shadow radius $r_{sh}=r_{sh}(M,\hat{\alpha})$, evaluated at the photon sphere~\cite{Ladino_2023, Antoniou2023, Vagnozzi_2023, ahmed2026shadow, Karthik2025, Zare_2026, Walia_2024, Lemos_2024, G_mez_2024, Tsukamoto_2024, Khodadi_2022}:
\begin{equation}
r_{sh} = b_c  = \frac{r_{ph}}{\sqrt{f(r_{ph})}} = \frac{r_{ph}^3}{\sqrt{r_{ph}^4 - 2M r_{ph}^3 + M^2 \hat{\alpha}}}.
\label{eq:shadow_radius}
\end{equation}
Due to the spherical symmetry of the qOS spacetime, $r_{sh}$ forms a perfect circle regardless of the observer's inclination $\theta_0$.
The shadow's optical properties depend strongly on $\hat{\alpha}$, revealing three distinct physical regimes based on the existence of the photon sphere \cite{Paul_2020,Tsukamoto_2021,Shaikh_2019,Chen2024}:

\begin{itemize}
\item \textbf{qOS black hole} ($\hat{\alpha} \leq \frac{27}{16} M^2$): Spacetime features an event horizon and a single unstable photon sphere ($r_{ph}$). The critical impact parameter $b_c$ marks the exact shadow boundary: photons with $b < b_c$ are captured by the horizon, forming a standard dark central shadow, while those with $b > b_c$ escape to infinity. Here, the bending angle diverges only as light rays approach the photon sphere from the exterior ($b \to b_c$ from $b > b_c$).

\item \textbf{qOS naked singularity with photon and anti-photon spheres} \cite{Paul_2020,Tsukamoto_2021,Shaikh_2019,Chen2024} ($\frac{27}{16} M^2 < \hat{\alpha} \leq \frac{729}{256} M^2$): The event horizon vanishes, leaving an unstable outer photon sphere ($r_{ph}$) and a stable inner anti-photon sphere ($r_{aph}$). At the upper limit ($\hat{\alpha} = \frac{729}{256} M^2$), both spheres merge into a single inflection point of the effective potential. Although $b_c$ remains well-defined, photons with $b < b_c$ are no longer captured; they enter the photon sphere and rebound off the infinitely high potential barrier near the singularity, causing the bending angle to diverge from both sides ($b > b_c$ and $b < b_c$).

\item \textbf{qOS naked singularity without photon spheres} ($\hat{\alpha} > \frac{729}{256} M^2$): Equations for circular photon orbits yield no real positive roots. Thus, the effective potential lacks a local maximum, leaving $b_c$ entirely undefined. No shadow boundary forms; all light rays are deflected or reflected back to infinity by the central potential barrier.
\end{itemize}

As $\hat{\alpha}\to0$, the Schwarzschild values $r_{ph}=3M$ and $r_{sh}=3\sqrt{3}M\approx 5.1965M$ are recovered. The characteristic radii $r_h$, $r_{ph}$, $r_{sh}$, and $r_{ISCO}$ describe test-particle motion in qOS spacetime. Although traditionally expressed in $(M,\hat{\alpha})$, using $M(S,\hat{\alpha})$ from Eq.~\eqref{eq:Mass_Entropy} maps them into entropy space, linking particle motion to black hole thermodynamics. As shown in Fig.~\ref{ShadowProfiles1} and Table~\ref{tab:unified-observables}, the characteristic radii decrease with $\hat{\alpha}$ in parameter space, but increase monotonically with entropy $S$ in the black-hole regime. We also provide the Reissner-Nordström black hole values for comparison, and distinguish thermodynamically stable and unstable black holes from the naked singularity regime, following Eq.~\eqref{eq:branches}.

\subsection{Ray Tracing with Thin Accretion Disks}


\begin{figure*}[h]
 \includegraphics[width=0.99\linewidth]{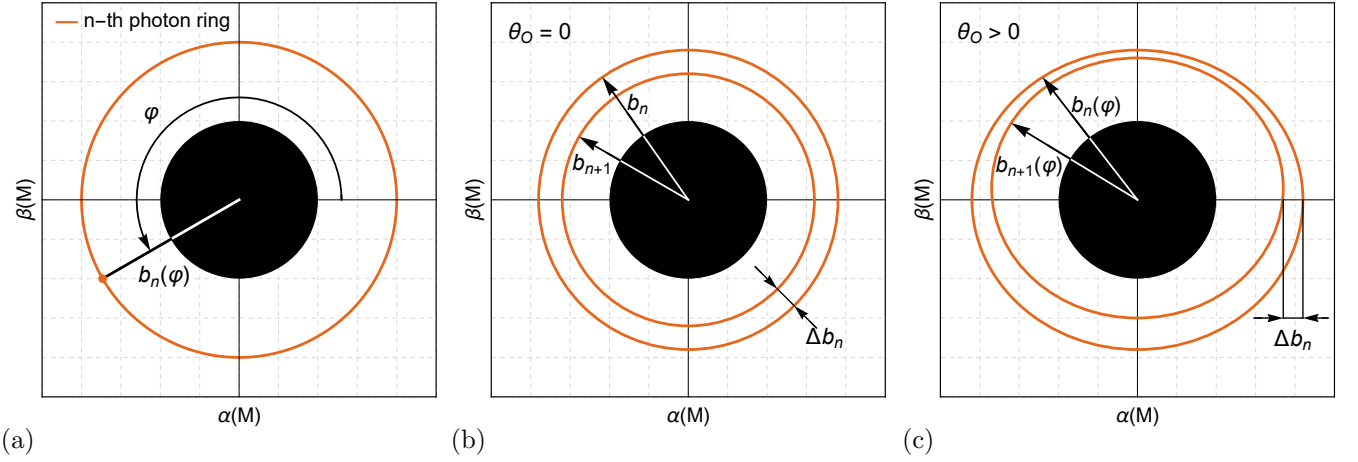}\vspace{-0.3cm}
 (a)\hspace{5.5cm}(b)\hspace{5.5cm}(c)\\
\vspace{-0.1cm}
\caption{Impact parameter configurations on the observer's screen with celestial coordinates $\alpha(M)$ and $\beta(M)$, showing the black hole and photon rings. (a) Impact parameter $b_n(\varphi)$ for the $n$-th ring. (b) For $\theta_0=0^\circ$, the rings are concentric, with constant gap $\Delta b_n$ between $b_n$ and $b_{n+1}$. (c) For $\theta_0>0^\circ$, the rings deform and $b_n(\varphi)$ depends on $\varphi$, while $\Delta b_n$ is conserved at $\varphi\in\{0^\circ,180^\circ\}$, making it optimal to measure $\Delta_n$ defined in Eq.~\eqref{deltan}.  Adapted from \cite{Aratore2024}.}

\label{photonringangles}
\end{figure*}


\begin{figure*}
    \centering
    \includegraphics[width=1\linewidth]{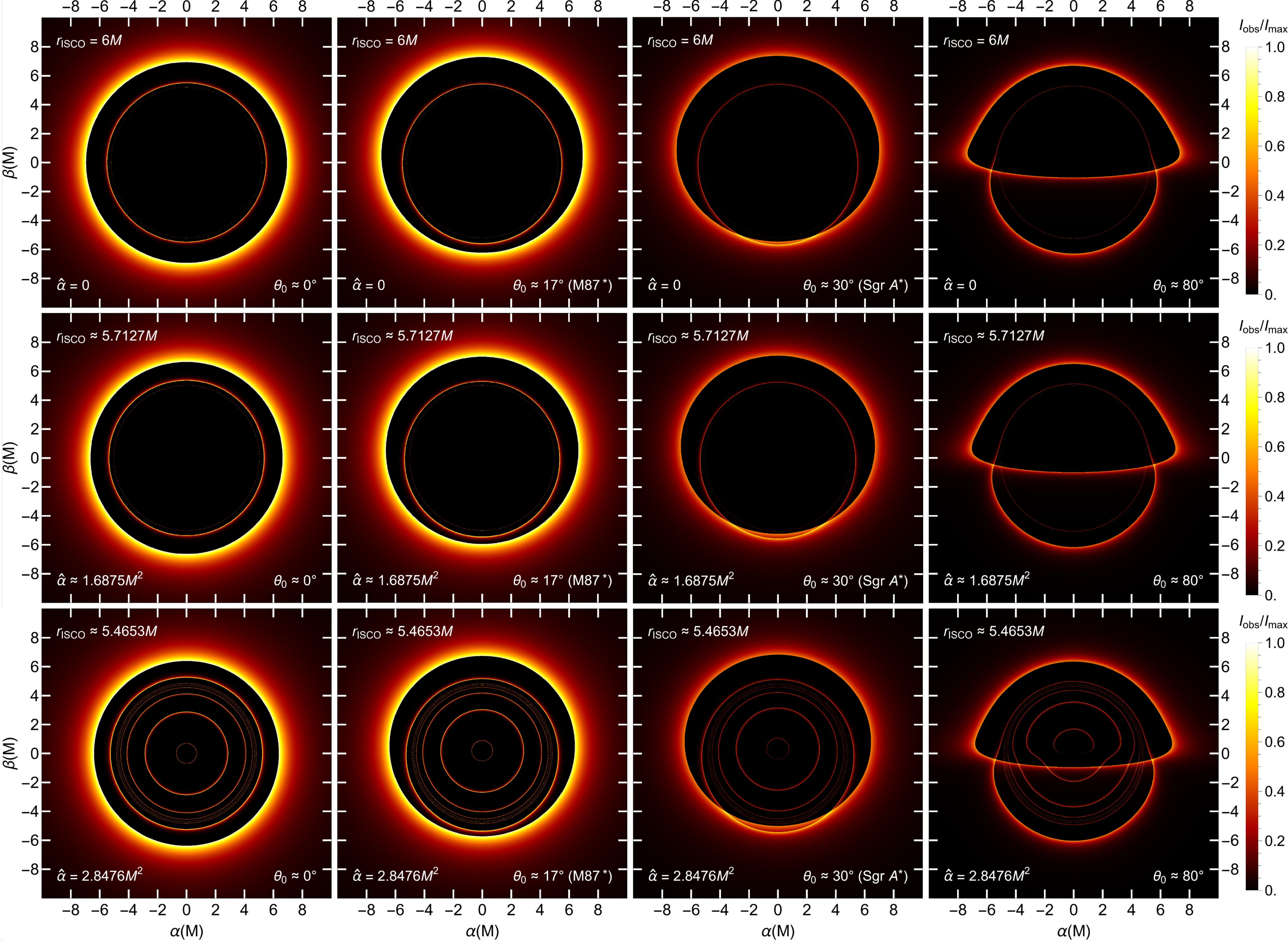}
\caption{Shadow images generated using \textbf{Profile~1} from $r_{ISCO}$ in Eq.~\eqref{profile1} at various observer inclinations, from left to right: $\theta_0 = 0^\circ$, $17^\circ$ for M87*, $30^\circ$ for Sgr~A*, and $80^\circ$. Top row: Schwarzschild black holes. Center row: Extremal qOS black holes. Bottom row: qOS naked singularities.}
    \label{fig:rt1}
\end{figure*}
\begin{figure*}
    \centering
    \includegraphics[width=1\linewidth]{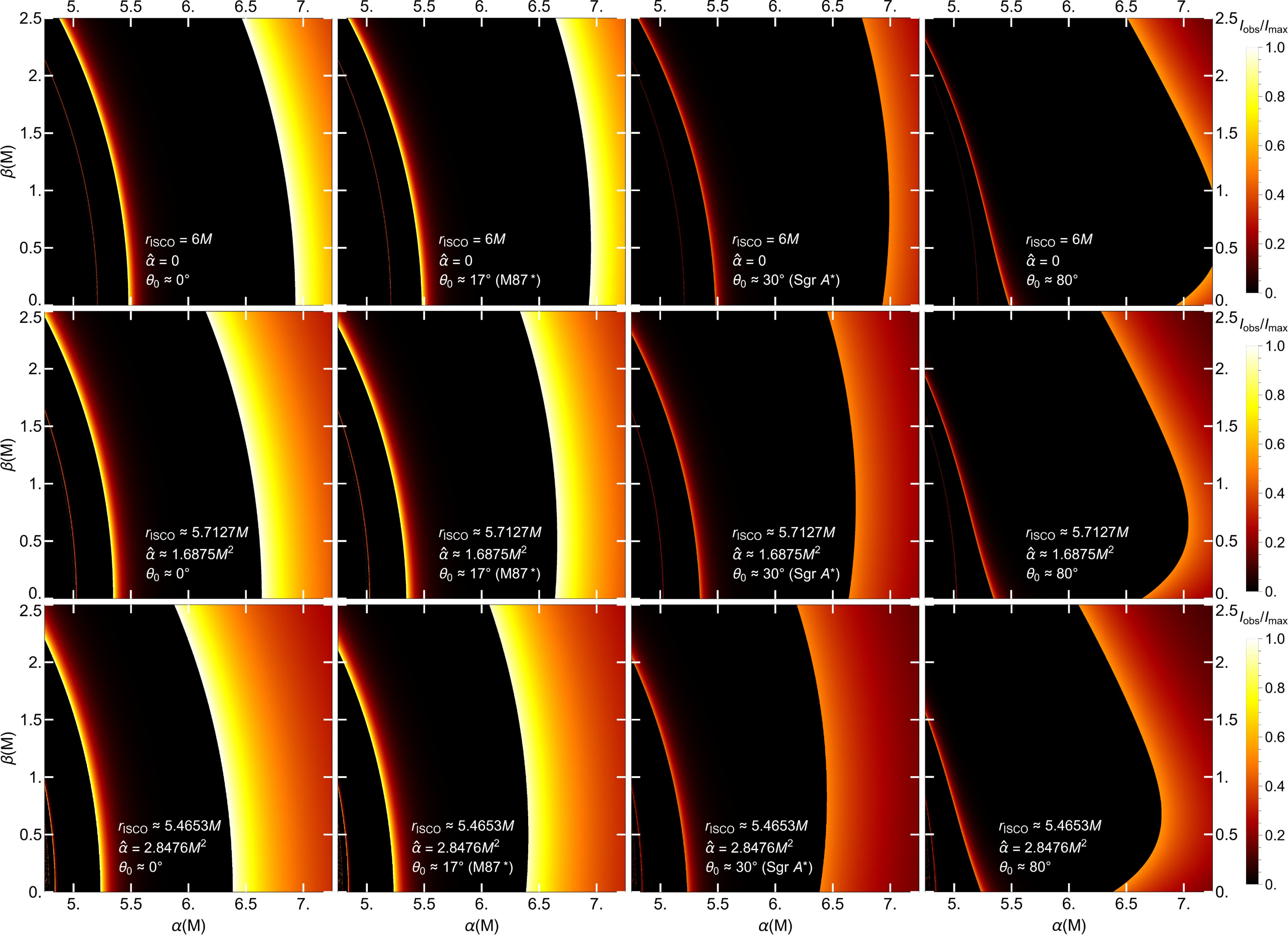}
\caption{Zoomed-in view of the direct image $b_0$ and photon rings $b_{n \geq 1}$ for the configurations in Fig.~\ref{fig:rt1}.}
    \label{fig:rt2}
\end{figure*}
\begin{figure*}
    \centering
    \includegraphics[width=1\linewidth]{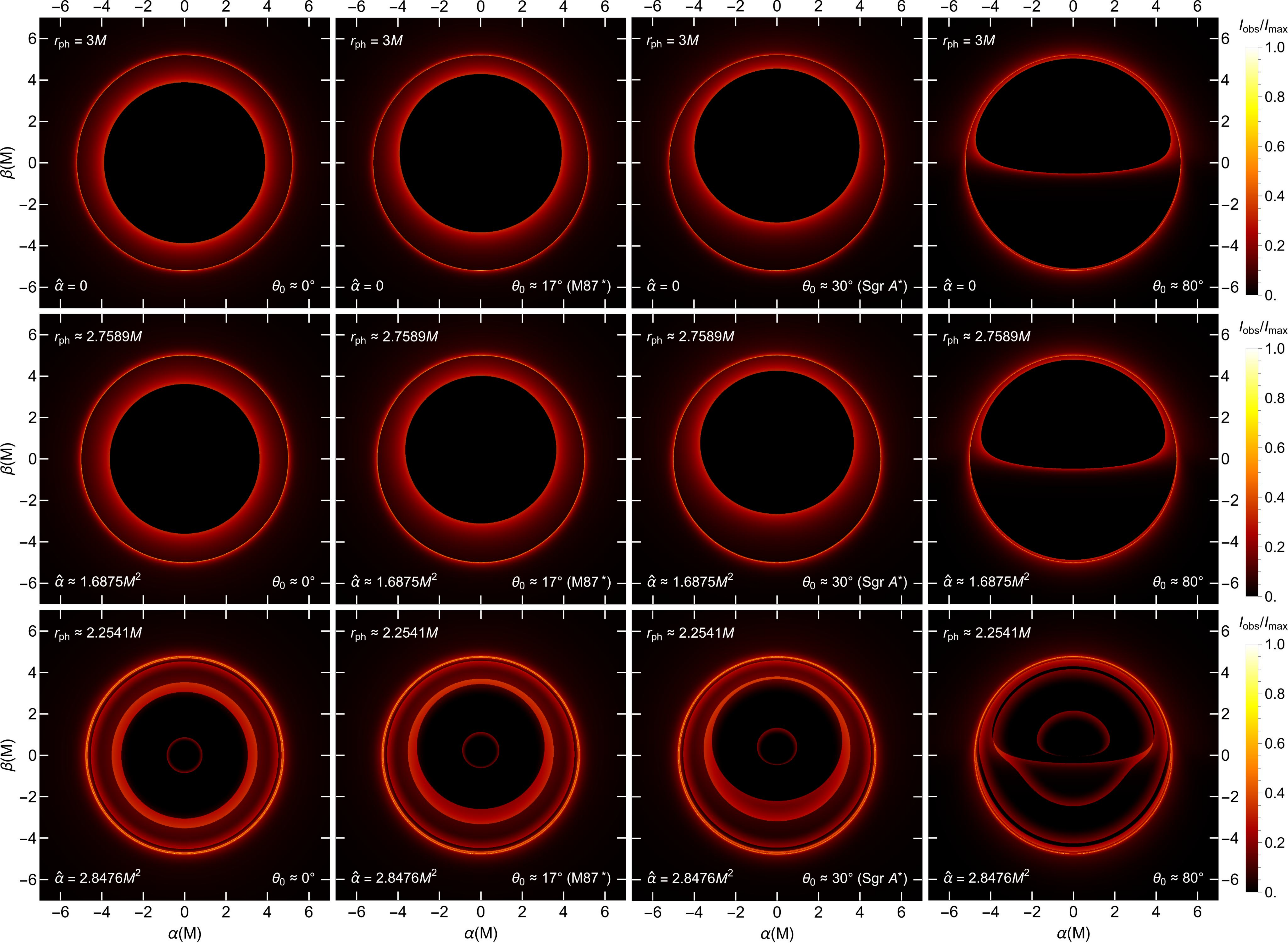}
    \caption{Shadow images generated using \textbf{Profile~2} from $r_{ph}$ in Eq.~\eqref{profile2} at various observer inclinations, from left to right: $\theta_0 = 0^\circ$, $17^\circ$ for M87*, $30^\circ$ for Sgr~A*, and $80^\circ$. Top row: Schwarzschild black holes. Center row: Extremal qOS black holes. Bottom row: qOS naked singularities.}
    \label{fig:rt3}
\end{figure*}
\begin{figure*}
    \centering
    \includegraphics[width=1\linewidth]{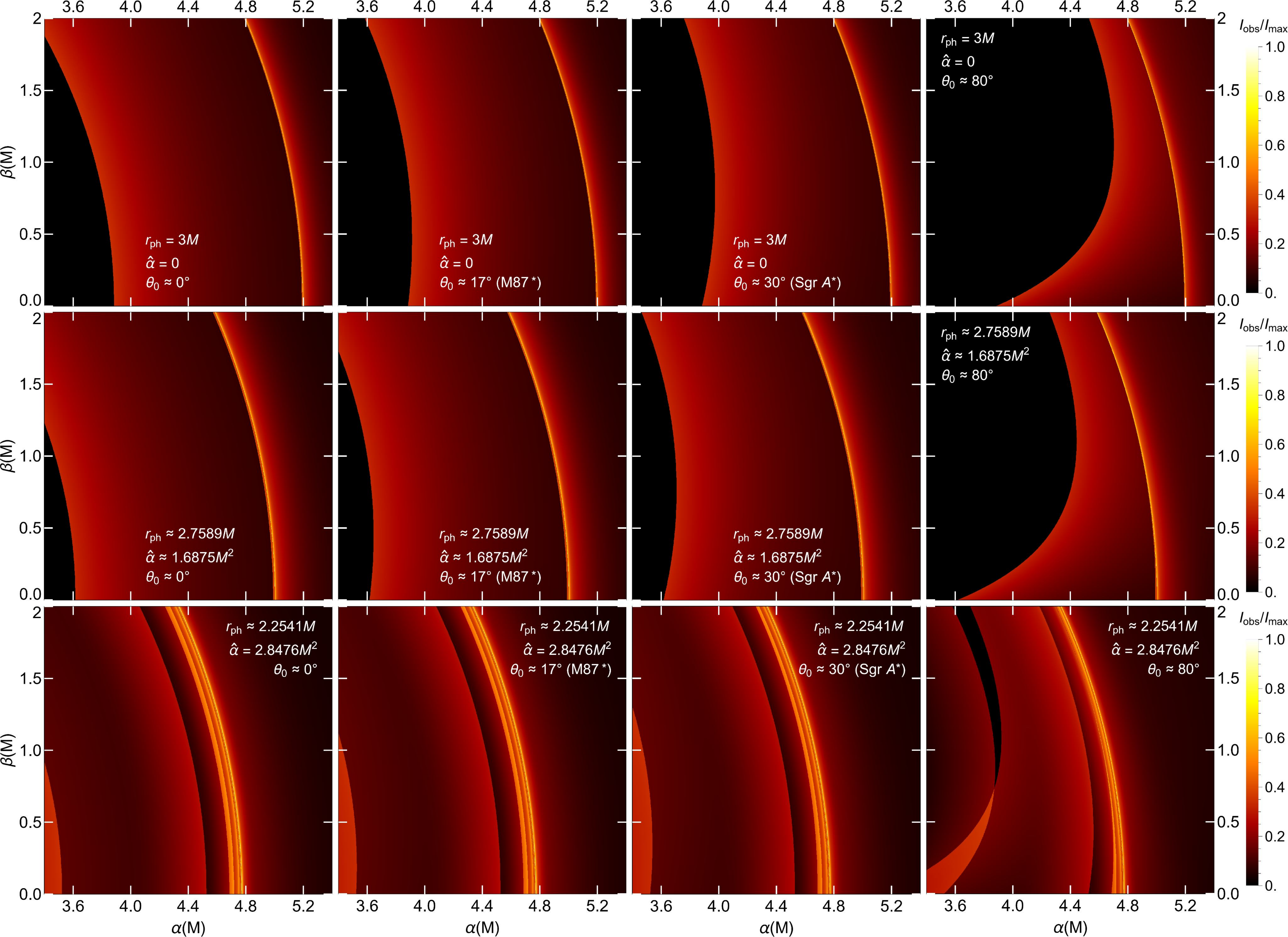}
\caption{Zoomed-in view of the direct image $b_0$ and photon rings $b_{n \geq 1}$ for the configurations in Fig.~\ref{fig:rt3}.}
    \label{fig:rt4}
\end{figure*}
\begin{figure*}
    \centering
    \includegraphics[width=1\linewidth]{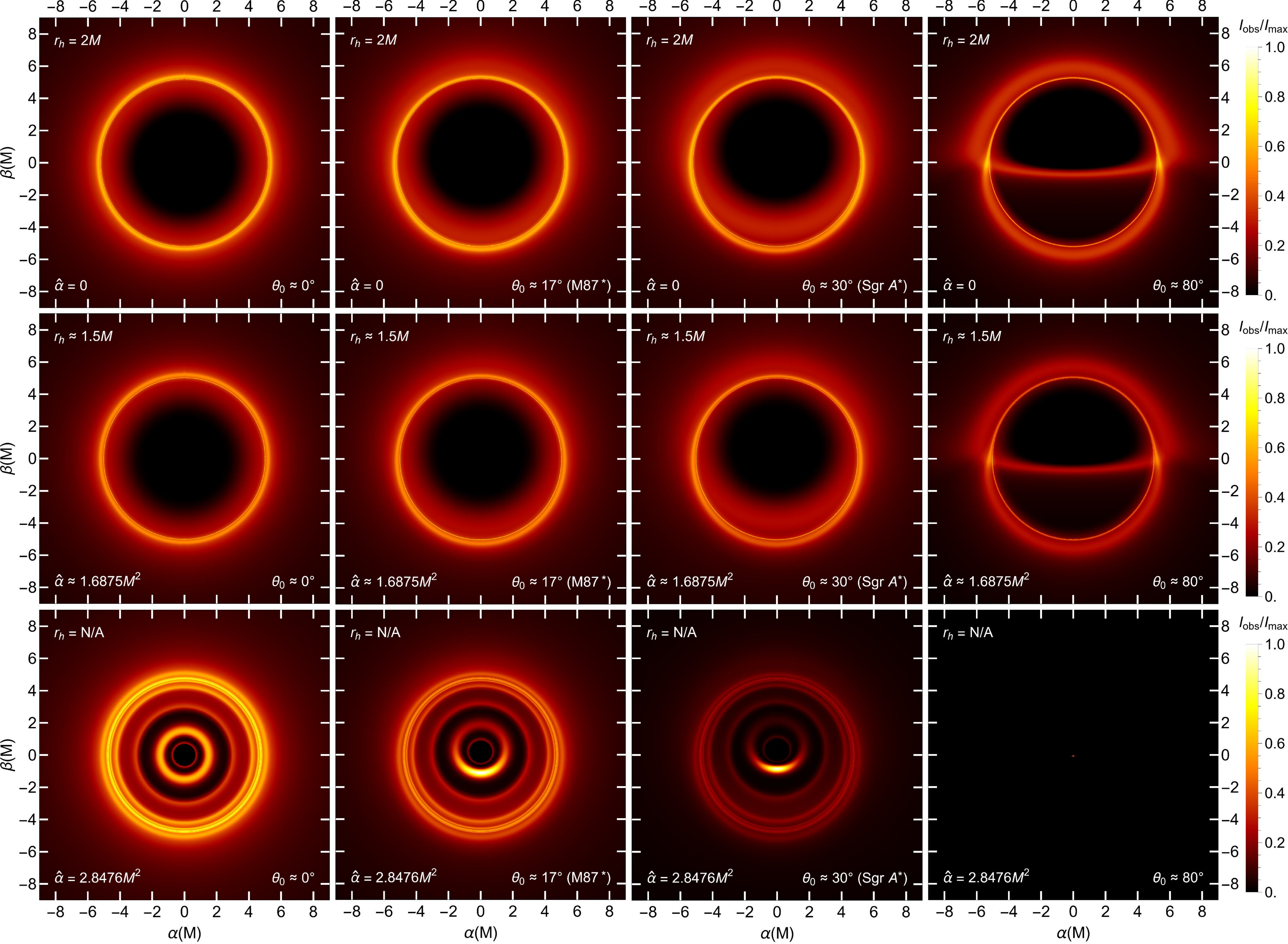}
\caption{Shadow images generated using \textbf{Profile~3} from $r_{h}$ in Eq.~\eqref{profile3} at various observer inclinations, from left to right: $\theta_0 = 0^\circ$, $17^\circ$ for M87*, $30^\circ$ for Sgr~A*, and $80^\circ$. Top row: Schwarzschild black holes. Center row: Extremal qOS black holes. Bottom row: qOS naked singularities (assuming $r = 0.05M$ in place of $r_h$ in Eq.~\eqref{profile3}}).
    \label{fig:rt5}
\end{figure*}
\begin{figure*}
    \centering
    \includegraphics[width=1\linewidth]{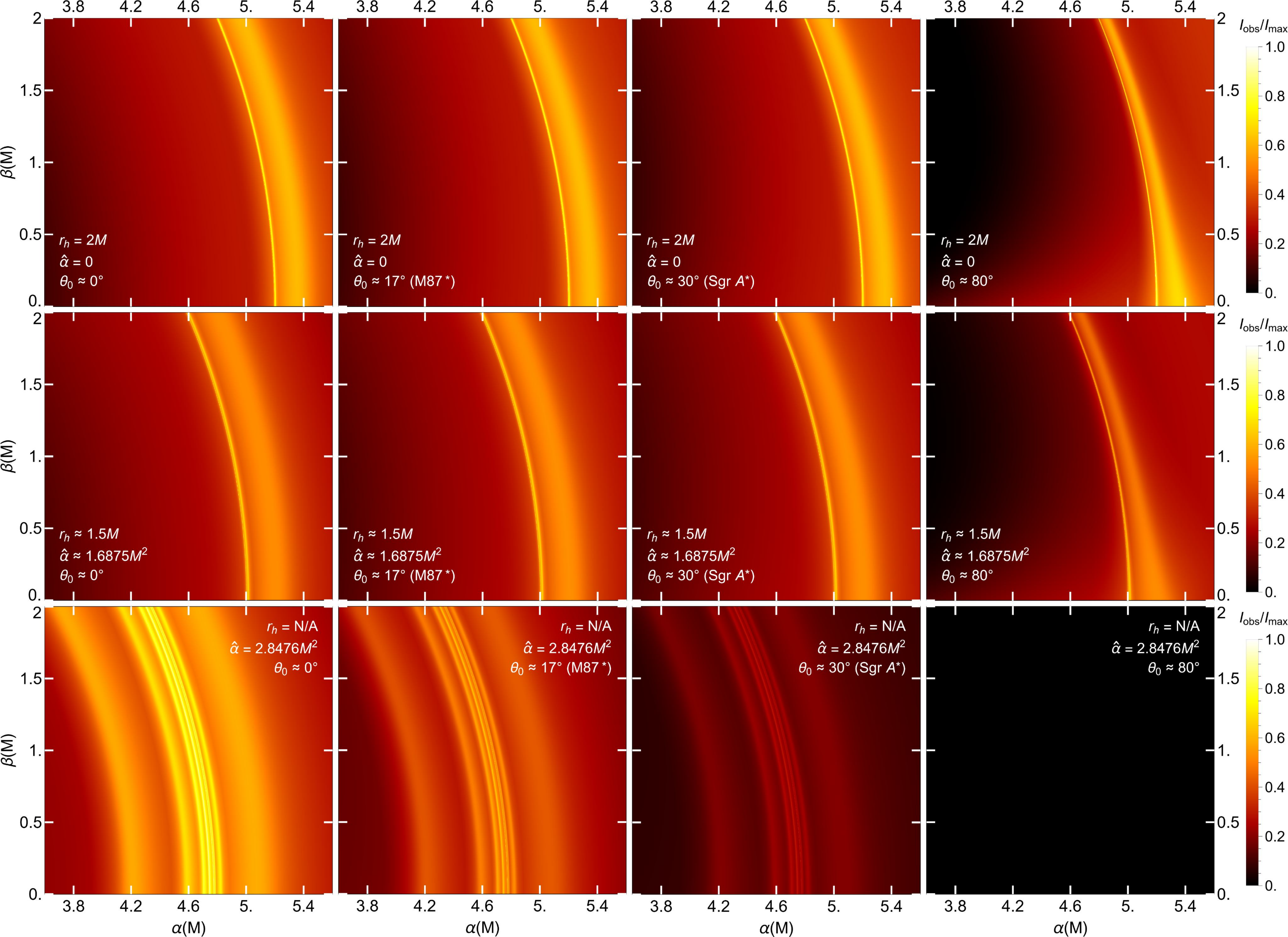}
\caption{Zoomed-in view of the direct image $b_0$ and photon rings $b_{n \geq 1}$ for the configurations in Fig.~\ref{fig:rt5}.}
    \label{fig:rt6}
\end{figure*}

The critical shadow boundary is accompanied by photon rings: thin, strongly lensed structures formed by photons completing one or more half-orbits near the photon sphere before escaping to infinity ($b>b_c$). Since $b\sim b_c$, these nested rings exponentially accumulate near the shadow edge.  As illustrated in Fig. \ref{photonringangles}, the configuration of these photon rings on the observer's screen depends on the observer's inclination $\theta_0$, with the impact parameters $b_n(\varphi)$ exhibiting a clear angular dependence and deforming from perfect circles when $\theta_0 > 0$ \cite{Aratore2024}.

To study photon rings and shadows in qOS spacetime, we model a geometrically and optically thin accretion disk in the equatorial plane ($\theta=\pi/2$), neglecting absorption and reflection \cite{Wang_2022,Yang_2023,Wang_2023,Chen2026,Zeng2023,Olmo_2023,gralla2019black,Gong_2026}. The radiative properties of thin accretion disks around qOS black holes, including the effect of an external magnetic field, have also been studied~\cite{qos5,idrissov2026magnetic}. The disk radiates isotropically in its local rest frame and is observed by a distant static observer at $r_0$ with inclination $\theta_0$. We use a static toy emissivity profile, rather than a dynamical accretion flow, to isolate geometric lensing, omitting  general relativistic magnetohydrodynamic (GRMHD) effects.  Along a photon geodesic, Liouville's theorem conserves $I_\nu/\nu^3$ \cite{Zeng2023,Olmo_2023,Gong_2026}, giving $I_{o,\nu_o}=g_r^3I_{e,\nu_e}$, where $g_r=\nu_o/\nu_e=\sqrt{f(r)}$ is the gravitational redshift factor. Integrating over observed frequencies, the total observed intensity from a single disk intersection is
\begin{equation}
    I_{\text{obs}}(r) = \int I_{o, \nu_o} \, d\nu_o = \int g_{r}^4 I_{e, \nu_e} \, d\nu_e = f(r)^2 I_{\text{em}}(r),
\end{equation}
where $I_{\text{em}}(r) = \int I_{e, \nu_e} \, d\nu_e$ corresponds to the total emission intensity in the rest frame of the accretion disk \cite{Wang_2022, Yang_2023, Wang_2023, Chen2026, Zeng2023}. Strong lensing allows multiple disk crossings. The transfer function $r_m(b)$ maps $b$ to the $m$-th intersection radius, and $I_{\text{obs}}(b)$ sums the redshifted emissions from all crossings
\begin{equation}
    I_{\text{obs}}(b) = \sum_{m} f(r)^2 I_{\text{em}}(r) \Big|_{r=r_m(b)}.
\end{equation}

To simulate the shadow images, we adopt three toy model emission profiles $I_{\text{em}}(r)$ ($M=1$) \cite{Wang_2022, Yang_2023, Wang_2023, Chen2026, Zeng2023}:

\begin{itemize}
    \item \textbf{Profile 1:} second-order decay starting from $r_{ISCO}$,
    \begin{equation}
        I_{\text{em}}(r) := \begin{cases} 
        I_{max} \left[ \frac{1}{r-(r_{ISCO}-1)} \right]^2, & r > r_{ISCO}, \\ 
        0, & r \leq r_{ISCO}. 
        \end{cases}
        \label{profile1}
    \end{equation}

    \item \textbf{Profile 2:} cubic decay starting from $r_{ph}$,
    \begin{equation}
        I_{\text{em}}(r) := \begin{cases} 
        I_{max} \left[ \frac{1}{r-(r_{ph}-1)} \right]^3, & r > r_{ph}, \\ 
        0, & r \leq r_{ph}. 
        \end{cases}
        \label{profile2}
    \end{equation}

    \item \textbf{Profile 3:} moderate decay starting from $r_{h}$,
    \begin{equation}
        I_{\text{em}}(r) := \begin{cases} 
        I_{max} \frac{\pi/2 - \arctan[r-(r_{ISCO}-1)]}{\pi/2 - \arctan[r_h-(r_{ISCO}-1)]}, & r > r_h, \\ 
        0, & r \leq r_h.
        \end{cases}
        \label{profile3}
    \end{equation}
\end{itemize}

Here, $I_{max}$ is the maximum emission intensity. Each screen pixel $(\alpha,\beta)$ defines the initial position $x^\mu_0$ and momentum $p_{\mu,0}$ for solving the null geodesics Eqs.~\eqref{eqmotion1}--\eqref{eqmotion3}. Trajectories are integrated backward, recording each equatorial crossing ($\theta=\pi/2$) to accumulate the emission. For black holes, crossings occur only outside the horizon, whereas horizonless naked singularities allow rays to traverse the central region, producing distinct central luminous structures instead of a dark inner shadow \cite{Chen2024,Fauzi_2025,deliyski2025observing,Rosa_2024}. Integration ends when rays escape to large radii, reach the horizon in black holes, or approach $r\sim0$ in naked singularities. The three emission models yield images indexed by $n\ge0$, counting photon half-orbits \cite{Aratore2024,Tsupko_2022}. As shown in Figs.~\ref{fig:rt1}--\ref{fig:rt6}, $n=0$ is the direct image, $n=1$ the first photon ring, and $n\ge2$ higher-order rings that asymptotically approach the photon sphere.
Figures~\ref{fig:rt1}--\ref{fig:rt6} illustrate these features for \textbf{Profile~1} (from $r_{\text{ISCO}}$), \textbf{Profile~2} (from $r_{\text{ph}}$), and \textbf{Profile~3} (from $r_{\text{h}}$), alongside their respective zoomed-in views. Each shadow-image set considers four inclinations, $\theta_0=0^\circ$, $17^\circ$ for M87* \cite{EHTM87}, $30^\circ$ for Sgr~A* \cite{EHTSgrA}, and $80^\circ$, for Schwarzschild, extremal qOS black holes, and qOS naked singularities. The parameter $\hat{\alpha}$ affects all impact parameters similarly to the characteristic radii of test-particle motion. As $\hat{\alpha}$ increases, the impact parameters $b_n$ decrease, reducing the shadow size, in agreement with the reduction of the photon capture cross section with increasing $\hat{\alpha}$ found in~\cite{idrissov2026magnetic}. For the qOS naked singularity, we choose $\hat{\alpha}$ such that both photon and anti-photon spheres exist \cite{Paul_2020,Tsukamoto_2021,Shaikh_2019,Chen2024}. Without an event horizon, \textbf{Profile~3} assumes emission near the origin ($r=0.05M$). Consequently, qOS black holes and naked singularities exhibit markedly different central images across all profiles \cite{Chen2024,Fauzi_2025,deliyski2025observing,Rosa_2024}.

In the black hole regime, light with $b<b_c$ is captured by the event horizon, producing a dark inner shadow. In qOS naked singularities with photon and anti-photon spheres, the infinite repulsive core prevents photon trapping, strongly deflecting light back to the observer and producing bright central rings from higher-order disk images \cite{Fauzi_2025,deliyski2025observing}. This behavior is similar to that obtained in the shadow images of other horizonless spacetimes or Exotic Compact Objects, such as gravastars or wormholes \cite{Fauzi_2025,Rosa_2024,Chernov2023}. Notably, the emission from these inner rings can be highly intense, potentially constituting the brightest region of the entire image. This effect is strikingly evident for \textbf{Profile~3} (Figs.~\ref{fig:rt5} and \ref{fig:rt6}): as $\theta_0$ increases, strong gravitational lensing forces lines of sight to spiral deeply and cross the equatorial disk multiple times. Summing the extreme emissions from these consecutive inner-region crossings produces a localized intensity peak. By $\theta_0 = 80^\circ$, this central radiation forms a compact, overwhelmingly luminous region that visually washes out the rest of the image due to the $I_{\text{obs}}/I_{\text{max}}$ normalization. Furthermore, Fig.~\ref{photonringangles} shows that $\Delta b_n=b_n-b_{n+1}$ is constant for $\theta_0=0^\circ$, where the rings are concentric and independent of $\varphi$. For $\theta_0>0^\circ$, the rings deform and $b_n(\varphi)$ and $b_{n+1}(\varphi)$ depend on $\varphi$.

Despite this deformation, the gap $\Delta b_n$ remains effectively conserved at $\varphi \in \{0^\circ, 180^\circ\}$~\cite{Aratore2024, Tsupko_2022}. For this reason, the zoomed-in views in Figs.~\ref{fig:rt2}, \ref{fig:rt4}, and \ref{fig:rt6} focus on the region around $\varphi = 0^\circ$ to highlight these conservation and deformation effects along the photon rings. Regardless of $\theta_0$, $\Delta b_n$ remains constant along the horizontal axis ($\varphi=0^\circ$) for Schwarzschild, extremal qOS black holes, and qOS naked singularities. In Sec.~\ref{sec3.4}, we characterize these gaps using $\Delta_n$, which provides measurable observables to distinguish qOS black holes, naked singularities, and other compact objects. 

Using \textbf{Profile~1} from Eq.~\eqref{profile1} in Figs.~\ref{fig:rt1} and \ref{fig:rt2}, emission begins at $r_{ISCO}$, where gravity is weaker. Since $I_{\text{obs}}\propto g_r^4$, photons lose less energy, yielding the highest luminosity among the three models \cite{Zeng2023}. The direct image is a redshifted and distorted source profile \cite{gralla2019black}. Three intensity peaks appear: the direct image, the first photon ring, and a third peak from overlapping higher-order rings. These rings have impact parameters $b_n$ and exponentially decreasing thickness, accumulating near the critical boundary $b_c$.
In the zoomed-in view, individual photon rings up to $n=6$ can be distinguished, with their impact parameters measured numerically, as shown in Table~\ref{tab:combined-model-I}. In contrast, \textbf{Profile~2} from Eq.~\eqref{profile2} in Figs.~\ref{fig:rt3} and \ref{fig:rt4} assumes emission begins at $r_{ph}$. Here, the observed peak values are notably attenuated by the redshift effect. Additionally, this model produces smaller shadows than \textbf{Profile~1}, placing the direct image inside the photon rings where their contributions superimpose and become difficult to distinguish. Furthermore, \textbf{Profile~3} from Eq.~\eqref{profile3} in Figs.~\ref{fig:rt5} and \ref{fig:rt6} assumes emission starts at $r_{h}$, where intensity reaches its maximum before gradually decaying outwards. This attenuation broadens the intensity peak of the first ring, allowing it to contribute more to the observed flux. The brightest region is a combination of the first ring and higher-order photon rings; however, they are observationally difficult to distinguish due to their overlap. Beyond this central bright region, direct emission still makes the main overall contribution to the flux. Interestingly, this blurred and superimposed appearance closely resembles the actual EHT shadow images \cite{EHTSgrA,EHTM87}.

Ultimately, these toy models show that observed intensity, ring structure, and overall brightness distribution depend heavily on both emission properties, such as disk boundaries and redshift effects, and the underlying spacetime geometry dictating light trajectories. While the accretion disk’s emission profile governs shadow intensity and appearance, making shadow size greatly model-dependent, assuming a fixed distance to the source and an observer position, the spacetime geometry strictly dictates photon ring positioning. Consequently, higher-order photon rings ($n \ge 1$) enable precise probes of underlying gravitational fields and possible quantum corrections. Building on this framework, the Black Hole Explorer (BHEX)~\cite{Johnson} aims to resolve the $n=1$ photon ring, which could provide a stringent observational test of the qOS spacetime model. By constraining the ring's boundaries $b_{1}^{\text{in}} \in (5.34570M, 5.47892M)$ and $b_{1}^{\text{out}} \in (6.09538M, 6.16755M)$, this observation could serve as a direct benchmark to distinguish the qOS model from General Relativity and alternative gravitational theories.

\subsection{Semi-Analytical Photon Rings}\label{sec3.3}

\begin{table*}[t]
\centering
\caption{Comparison of semi-analytical (with Eq.~(\ref{eq:bn_explicit})) and numerical ray tracing results: range of impact parameters $b_n$ and their values at the Davies point for Quantum Oppenheimer-Snyder black holes. The light source is the thin accretion disk with Profile~1 from Eq.~\eqref{profile1} where $r_S\in [r_{ISCO}, \infty)$.}
\label{tab:combined-model-I}
\scriptsize 
\setlength{\tabcolsep}{4pt} 
\renewcommand{\arraystretch}{1.2} 
\begin{tabular}{c|c|c|c|c|c}
\hline
 \multicolumn{2}{c|}{} & \multicolumn{2}{c|}{\textbf{$b_n$ Range}} & \multicolumn{2}{c}{\textbf{$b_n$ at Davies Point}} \\
 \multicolumn{2}{c|}{} & \multicolumn{2}{c|}{$[b_n(T=0), b_n(T=T_{Schw})]$} & \multicolumn{2}{c}{$b_{nm}=b_n(T=T_m)$} \\ \hline
\multicolumn{2}{c|}{\begin{tabular}{@{}c@{}} \textbf{} \\ \textbf{} \end{tabular}} & \begin{tabular}{@{}c@{}}\textbf{Parameter Space} \\ $b_n(M)$ vs $\hat{\alpha}(M^2)$\end{tabular} & \begin{tabular}{@{}c@{}}\textbf{Entropy Space} \\ $b_n(\sqrt{\hat{\alpha}})$ vs $S(\hat{\alpha})$\end{tabular} & \begin{tabular}{@{}c@{}}\textbf{Parameter Space} \\ $b_n(M)$ vs $\hat{\alpha}(M^2)$\end{tabular} & \begin{tabular}{@{}c@{}}\textbf{Entropy Space} \\ $b_n(\sqrt{\hat{\alpha}})$ vs $S(\hat{\alpha})$\end{tabular} \\ \hline

\multicolumn{2}{c|}{$b_n$} & $[b_n(\hat{\alpha}_{ext} = \frac{27 M^2}{16}), b_n(\hat{\alpha} = 0)]$ & $[b_n(S_{ext} = \frac{4 \pi \hat{\alpha}}{3}), b_n(S \to \infty))$ & $b_n(\hat{\alpha}_m = 1.2835 M^2)$ & $b_n(S_m = \frac{11+\sqrt{13}}{6} \pi \hat{\alpha})$ \\ \hline

\multirow{10}{*}{\rotatebox{90}{\textbf{Semi-Analytical}}} 
 & \multirow{2}{*}{$b_2$} & $b_2^{in} \in [5.02354M, 5.20795M]$ & $b_2^{in} \in [3.86713\sqrt{\hat{\alpha}}, \infty)$ & $b_{2m}^{in} \approx 5.07384M$ & $b_{2m}^{in} \approx 4.47855\sqrt{\hat{\alpha}}$ \\
 & & $b_2^{out} \in [5.05211M, 5.22743M]$ & $b_2^{out} \in [3.88912\sqrt{\hat{\alpha}}, \infty)$ & $b_{2m}^{out} \approx 5.09948M$ & $b_{2m}^{out} \approx 4.50119\sqrt{\hat{\alpha}}$ \\ 

 & \multirow{2}{*}{$b_3$} & $b_3^{in} \in [5.00358M, 5.19666M]$ & $b_3^{in} \in [3.85176\sqrt{\hat{\alpha}}, \infty)$ & $b_{3m}^{in} \approx 5.05706M$ & $b_{3m}^{in} \approx 4.46374\sqrt{\hat{\alpha}}$ \\
 & & $b_3^{out} \in [5.00550M, 5.19750M]$ & $b_3^{out} \in [3.85324\sqrt{\hat{\alpha}}, \infty)$ & $b_{3m}^{out} \approx 5.05855M$ & $b_{3m}^{out} \approx 4.46506\sqrt{\hat{\alpha}}$ \\ 

 & \multirow{2}{*}{$b_4$} & $b_4^{in} \in [5.00224M, 5.19617M]$ & $b_4^{in} \in [3.85072\sqrt{\hat{\alpha}}, \infty)$ & $b_{4m}^{in} \approx 5.05608M$ & $b_{4m}^{in} \approx 4.46288\sqrt{\hat{\alpha}}$ \\
 & & $b_4^{out} \in [5.00237M, 5.19621M]$ & $b_4^{out} \in [3.85082\sqrt{\hat{\alpha}}, \infty)$ & $b_{4m}^{out} \approx 5.05617M$ & $b_{4m}^{out} \approx 4.46296\sqrt{\hat{\alpha}}$ \\ 

 & \multirow{2}{*}{$b_5$} & $b_5^{in} \in [5.00215M, 5.19615M]$ & $b_5^{in} \in [3.85065\sqrt{\hat{\alpha}}, \infty)$ & $b_{5m}^{in} \approx 5.05602M$ & $b_{5m}^{in} \approx 4.46283\sqrt{\hat{\alpha}}$ \\
 & & $b_5^{out} \in [5.00216M, 5.19615M]$ & $b_5^{out} \in [3.85066\sqrt{\hat{\alpha}}, \infty)$ & $b_{5m}^{out} \approx 5.05603M$ & $b_{5m}^{out} \approx 4.46284\sqrt{\hat{\alpha}}$ \\ 

 & \multirow{2}{*}{$b_6$} & $b_6^{in} \in [5.00214M, 5.19615M]$ & $b_6^{in} \in [3.85065\sqrt{\hat{\alpha}}, \infty)$ & $b_{6m}^{in} \approx 5.05602M$ & $b_{6m}^{in} \approx 4.46283\sqrt{\hat{\alpha}}$ \\
 & & $b_6^{out} \in [5.00214M, 5.19615M]$ & $b_6^{out} \in [3.85065\sqrt{\hat{\alpha}}, \infty)$ & $b_{6m}^{out} \approx 5.05602M$ & $b_{6m}^{out} \approx 4.46283\sqrt{\hat{\alpha}}$ \\ \hline

\multirow{12}{*}{\rotatebox{90}{\textbf{Numerical Ray Tracing}}} 
 & \multirow{2}{*}{$b_0$} & $b_0^{in} \in [6.63907M, 6.93215M]$ & $b_0^{in} \in [5.11076\sqrt{\hat{\alpha}}, \infty)$ & $b_{0m}^{in} \approx 6.71559M$ & $b_{0m}^{in} \approx 5.92769 \sqrt{\hat{\alpha}}$ \\
 & & $b_0^{out} \rightarrow \infty$ & $b_0^{out} \rightarrow \infty$ & $b_{0m}^{out} \rightarrow \infty$ & $b_{0m}^{out} \rightarrow \infty$ \\

 & \multirow{2}{*}{$b_1$} & $b_1^{in} \in [5.34570M, 5.47892M]$ & $b_1^{in} \in [4.11512\sqrt{\hat{\alpha}}, \infty)$ & $b_{1m}^{in} \approx 5.38018M$ & $b_{1m}^{in} \approx 4.74896\sqrt{\hat{\alpha}}$ \\
 & & $b_1^{out} \in [6.09538M, 6.16755M]$ & $b_1^{out} \in [4.69223\sqrt{\hat{\alpha}}, \infty)$ & $b_{1m}^{out} \approx 6.11323M$ & $b_{1m}^{out} \approx 5.39600\sqrt{\hat{\alpha}}$ \\ 

 & \multirow{2}{*}{$b_2$} & $b_2^{in} \in [5.02376M, 5.20799M]$ & $b_2^{in} \in [3.86729\sqrt{\hat{\alpha}}, \infty)$ & $b_{2m}^{in} \approx 5.07396M$ & $b_{2m}^{in} \approx 4.47866\sqrt{\hat{\alpha}}$ \\
 & & $b_2^{out} \in [5.05371M, 5.22794M]$ & $b_2^{out} \in [3.89035\sqrt{\hat{\alpha}}, \infty)$ & $b_{2m}^{out} \approx 5.10054M$ & $b_{2m}^{out} \approx 4.50212\sqrt{\hat{\alpha}}$ \\ 

 & \multirow{2}{*}{$b_3$} & $b_3^{in} \in [5.00358M, 5.19666M]$ & $b_3^{in} \in [3.85176\sqrt{\hat{\alpha}}, \infty)$ & $b_{3m}^{in} \approx 5.05706M$ & $b_{3m}^{in} \approx 4.46375\sqrt{\hat{\alpha}}$ \\
 & & $b_3^{out} \in [5.00547M, 5.19751M]$ & $b_3^{out} \in [3.85321\sqrt{\hat{\alpha}}, \infty)$ & $b_{3m}^{out} \approx 5.05855M$ & $b_{3m}^{out} \approx 4.46506\sqrt{\hat{\alpha}}$ \\ 

 & \multirow{2}{*}{$b_4$} & $b_4^{in} \in [5.00224M, 5.19617M]$ & $b_4^{in} \in [3.85073\sqrt{\hat{\alpha}}, \infty)$ & $b_{4m}^{in} \approx 5.05608M$ & $b_{4m}^{in} \approx 4.46288\sqrt{\hat{\alpha}}$ \\
 & & $b_4^{out} \in [5.00236M, 5.19621M]$ & $b_4^{out} \in [3.85082\sqrt{\hat{\alpha}}, \infty)$ & $b_{4m}^{out} \approx 5.05616M$ & $b_{4m}^{out} \approx 4.46295\sqrt{\hat{\alpha}}$ \\ 

 & \multirow{2}{*}{$b_5$} & $b_5^{in} \in [5.00215M, 5.19615M]$ & $b_5^{in} \in [3.85066\sqrt{\hat{\alpha}}, \infty)$ & $b_{5m}^{in} \approx 5.05602M$ & $b_{5m}^{in} \approx 4.46283\sqrt{\hat{\alpha}}$ \\
 & & $b_5^{out} \in [5.00216M, 5.19615M]$ & $b_5^{out} \in [3.85066\sqrt{\hat{\alpha}}, \infty)$ & $b_{5m}^{out} \approx 5.05603M$ & $b_{5m}^{out} \approx 4.46284\sqrt{\hat{\alpha}}$ \\ 

 & \multirow{2}{*}{$b_6$} & $b_6^{in} \in [5.00214M, 5.19615M]$ & $b_6^{in} \in [3.85065\sqrt{\hat{\alpha}}, \infty)$ & $b_{6m}^{in} \approx 5.05602M$ & $b_{6m}^{in} \approx 4.46283\sqrt{\hat{\alpha}}$ \\
 & & $b_6^{out} \in [5.00214M, 5.19615M]$ & $b_6^{out} \in [3.85065\sqrt{\hat{\alpha}}, \infty)$ & $b_{6m}^{out} \approx 5.05602M$ & $b_{6m}^{out} \approx 4.46283\sqrt{\hat{\alpha}}$ \\ \hline

\end{tabular}
\end{table*}


\begin{figure*}[h]
\vspace{0.2cm}
\begin{minipage}[t]{0.475\linewidth}
\centering
\hspace{-7.2cm} \includegraphics[width=0.99\linewidth]{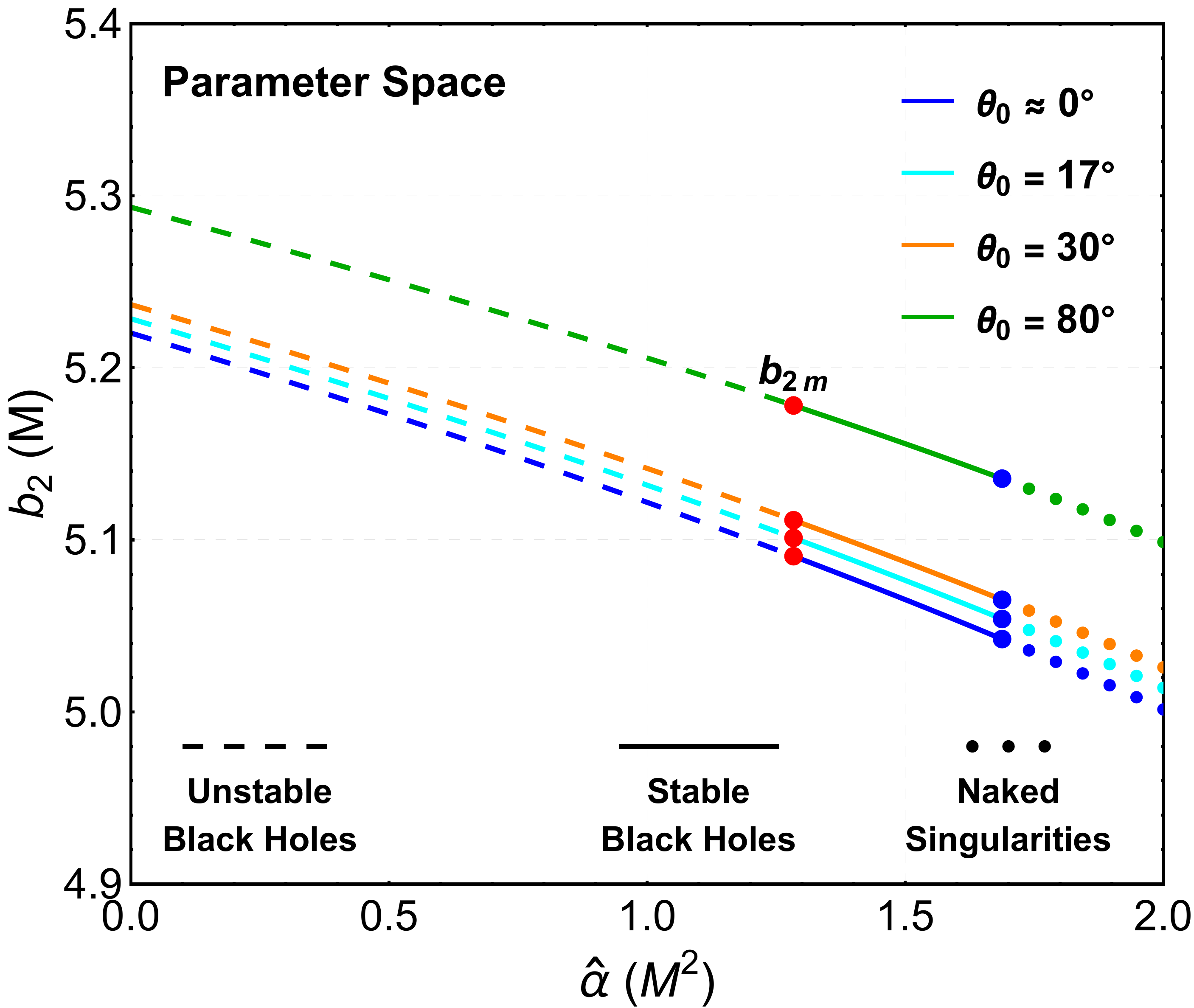}
\hspace{-8cm} (a)
\end{minipage}%
\hfill%
\begin{minipage}[t]{0.49\linewidth}
\centering
\hspace{-7.7cm} \includegraphics[width=0.975\linewidth]{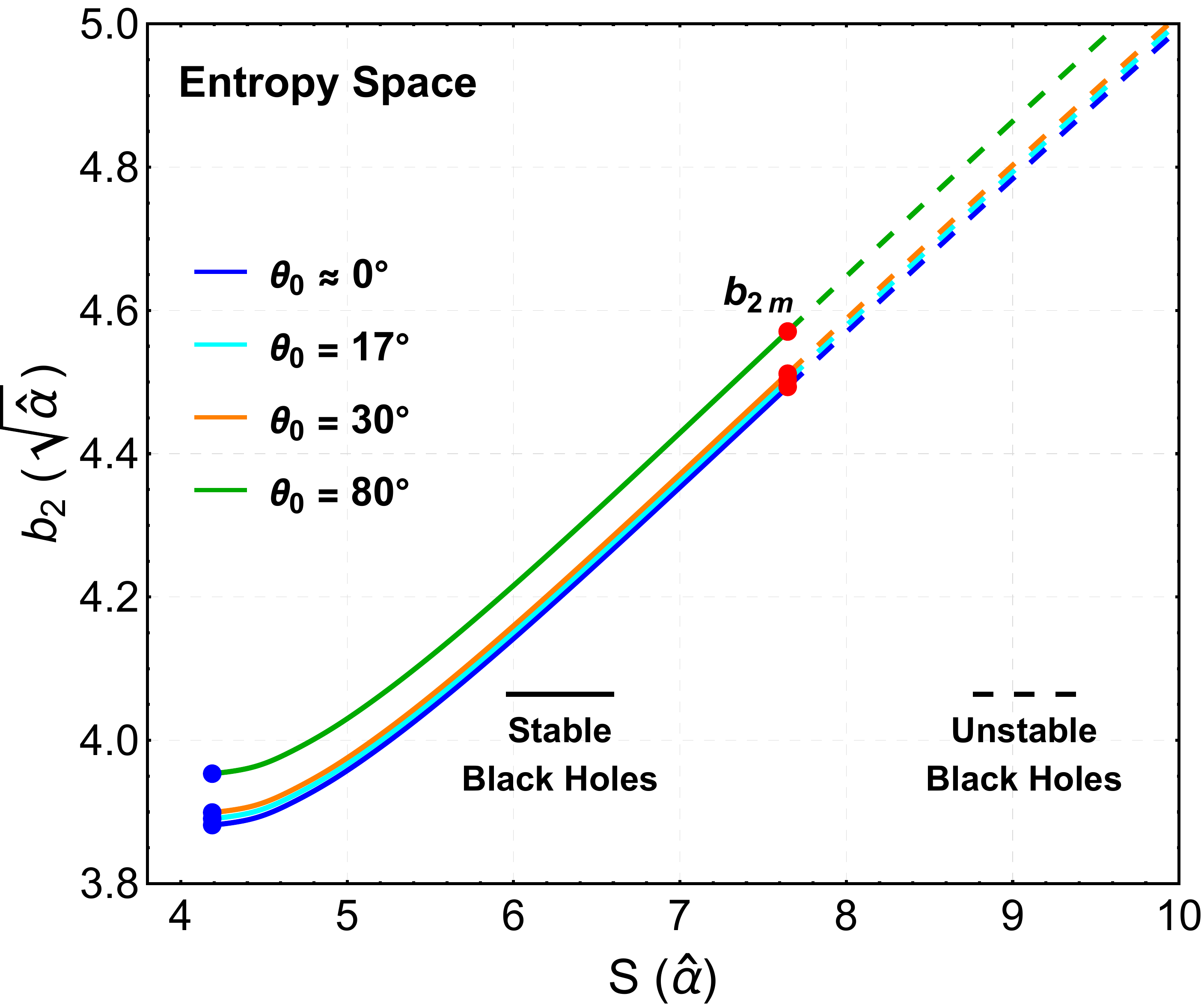}
\hspace{-8cm} (b)
\end{minipage}%
\hfill 
\begin{minipage}[t]{0.475\linewidth}
\centering
\hspace{-7.2cm} \includegraphics[width=0.99\linewidth]{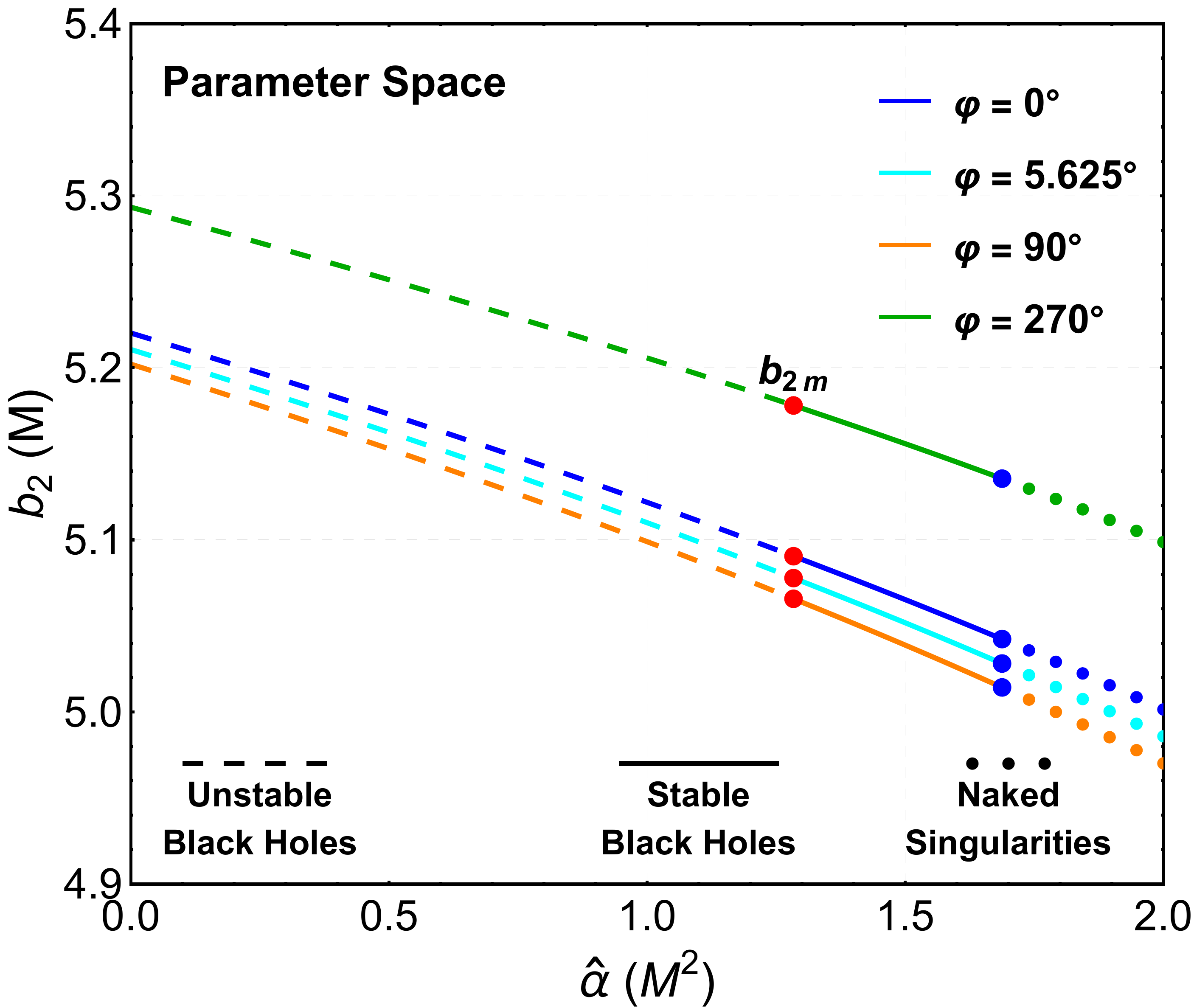}
\hspace{-8cm} (c)
\end{minipage}%
\hfill%
\begin{minipage}[t]{0.49\linewidth}
\centering
\hspace{-7.1cm} \includegraphics[width=0.98\linewidth]{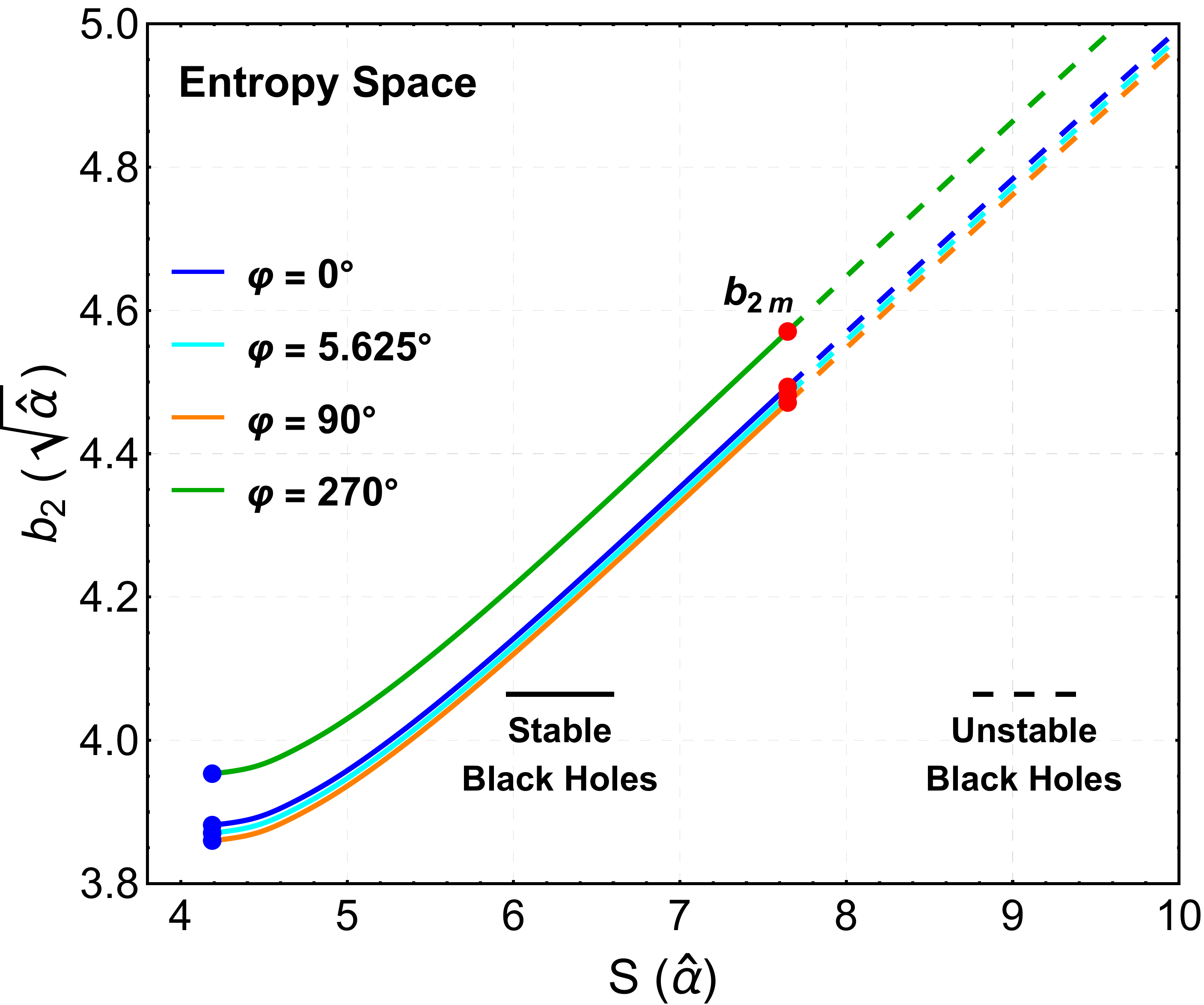}
\hspace{-8cm} (d)
\end{minipage}%
\hfill \
\vspace{0.02cm}
\caption{Behavior of the impact parameter $b_2$ for qOS spacetime. Panels (a,c) show $b_2$ versus $\hat{\alpha}$ for varying $\theta_0$ and screen polar angle $\varphi$, respectively, while (b,d) show the corresponding entropy-space results. Unstable and stable black holes and naked singularities are indicated. Red dots ($b_{2m}$) mark the Davies point, and blue dots the extremal case.}
\label{Shadows9}
\end{figure*}



\begin{table*}[t]
\centering
        \caption{Comparison of Semi-Analytical (S-A) results (Eq.~\eqref{eq:delta_n_explicit}) and Numerical Ray Tracing (NRT) for qOS black holes, showing the gap parameters $\Delta_n$ and their Davies-point values. The thin accretion disk follows Profile~1 (Eq.~\eqref{profile1}), with $r_S \in [r_{ISCO},\infty)$; $\Delta_n$ is computed from the inner boundaries $b_n^{in}$ of Table~\ref{tab:combined-model-I}, i.e., for emission at $r_S=r_{ISCO}$.}
        \label{tab:gap-parameters-unified}
        \resizebox{\linewidth}{!}{
        \renewcommand{\arraystretch}{1.1} 
        \begin{tabular}{c|c|c|c|c|c}
        \hline
        \textbf{} & \textbf{Method} & \textbf{$\Delta_n$ Range} & \textbf{$\Delta_n$ at Davies Point} & \textbf{Error Range} & \textbf{Error at Davies Point} \\
        \textbf{} & & $[\Delta_n^{\min}, \Delta_n^{\max}]$ for $T \in [0,T_{Schw}]$ & $\Delta_{nm}=\Delta_n(T=T_m)$ & $[E_r^{\min}, E_r^{\max}]$ for $T \in [0,T_{Schw}]$ & $E_{rm}=E_r(T=T_m)$ \\ \hline

        $\Delta_0$ & \textbf{NRT} & $\Delta_0 \in [0.19481, 0.20964]$ & $\Delta_{0m} \approx 0.19885$ & -- & -- \\ \hline
        
        $\Delta_1$ & \textbf{NRT} & $\Delta_1 \in [0.04945, 0.06022]$ & $\Delta_{1m} \approx 0.05692$ & -- & -- \\ \hline
        
        \multirow{2}{*}{$\Delta_2$} 
         & \textbf{S-A} & $\Delta_2 \in [0.00217, 0.00397]$ & $\Delta_{2m} \approx 0.00331$ & \multirow{2}{*}{$[0.46\%, 1.24\%]$} & \multirow{2}{*}{$0.60\%$} \\
         & \textbf{NRT}  & $\Delta_2 \in [0.00218, 0.00402]$ & $\Delta_{2m} \approx 0.00333$ & & \\ \hline
         
        \multirow{2}{*}{$\Delta_3$} 
         & \textbf{S-A} & $\Delta_3 \in [0.00009, 0.00027]$ & $\Delta_{3m} \approx 0.00019$ & \multirow{2}{*}{$[0.00\%, 0.00\%]$} & \multirow{2}{*}{$0.00\%$} \\
         & \textbf{NRT}  & $\Delta_3 \in [0.00009, 0.00027]$ & $\Delta_{3m} \approx 0.00019$ & & \\ \hline
         
        \multirow{2}{*}{$\Delta_4$} 
         & \textbf{S-A} & $\Delta_4 \in [0.00000, 0.00002]$ & $\Delta_{4m} \approx 0.00001$ & \multirow{2}{*}{$[0.00\%, 0.00\%]$} & \multirow{2}{*}{$0.00\%$} \\
         & \textbf{NRT}  & $\Delta_4 \in [0.00000, 0.00002]$ & $\Delta_{4m} \approx 0.00001$ & & \\ \hline
         
        \multirow{2}{*}{$\Delta_5$} 
         & \textbf{S-A} & $\Delta_5 \in [0.00000, 0.00000]$ & $\Delta_{5m} \approx 0.00000$ & \multirow{2}{*}{$[0.00\%, 0.00\%]$} & \multirow{2}{*}{$0.00\%$} \\
         & \textbf{NRT}  & $\Delta_5 \in [0.00000, 0.00000]$ & $\Delta_{5m} \approx 0.00000$ & & \\ \hline

        \end{tabular}
        
        } 
        \end{table*}

Following \cite{Aratore2024,Tsupko_2022}, we calculate the photon-ring impact parameters $b_n$ in a static, spherically symmetric, asymptotically flat spacetime by tracing light rays from a source at $r_S$ near the photon sphere $r_{ph}$. For a ray reaching an observer at $r_0$ after a minimum radius $R_t$, the total azimuthal deflection $\Delta\tilde{\phi}$ is given by \cite{Bozza_2007,Aratore2024}
\begin{equation}
    \Delta\tilde{\phi} = \int_{R_t}^{r_S} \frac{b}{r \sqrt{r^2 - b^2 f(r)}} \, dr + \int_{R_t}^{r_0} \frac{b}{r \sqrt{r^2 - b^2 f(r)}} \, dr.
    \label{eq:azimuthal_shift}
\end{equation}
As the turning point $R_t \to r_{ph}$, the impact parameter $b \to r_{sh}$. In this strong deflection limit, the integrals in Eq. (\ref{eq:azimuthal_shift}) diverge as the light ray completes multiple loops around the compact object. To treat this divergent regime analytically, it is convenient to parameterize $b$ as a small positive perturbation $\epsilon \ll 1$ relative to $r_{sh}$ via $b = r_{sh}(1 + \epsilon)$, where $\epsilon = (b - r_{sh})/r_{sh}$ \cite{Aratore2024}. As $\epsilon \to 0$, photons undergo multiple turns around the photon sphere, generating an infinite sequence of images indexed by $n \ge 0$. While this approximation is invalid for low-order images ($n=0,1$), it becomes highly accurate for higher-order images ($n \ge 2$). Focusing on these higher-order images, we consider an observer at infinity ($r_0 \to \infty$), where each image point on the observer's screen is described by its impact parameter $b$ and polar angle $\varphi$ (Fig. \ref{photonringangles}). Thus, regardless of whether $n$ is even or odd, the total azimuthal shift $\Delta\tilde{\phi}$ can be geometrically related to $\varphi$ and the observer's inclination angle $\theta_0$ via a unified relation \cite{Aratore2024, Tsupko_2022}
\begin{equation}
\Delta\tilde{\phi} = (n + 1)\pi - \arccos\left( \frac{\sin \varphi}{\sqrt{\sin^2 \varphi + \cot^2 \theta_0}} \right).
\label{eq:delta_phi_luminet}
\end{equation}
For a polar observer ($\theta_0 = 0$ or $\pi$), this relation simplifies to $\Delta\tilde{\phi} = \left( n + \frac{1}{2} \right)\pi$. Applying a logarithmic approximation \cite{Bozza_2007,Aratore2024} to the integrals in Eq.~(\ref{eq:azimuthal_shift}), and substituting Eq.~(\ref{eq:delta_phi_luminet}) allows us to invert the relation for the $n$-th order impact parameter $b_n$. In the qOS spacetime metric, described by the mass $M$ and quantum correction parameter $\hat{\alpha}$, since $r_{ph}=r_{ph}(M,\hat{\alpha})$, then $b_n = b_n(n, \varphi, r_S, r_0, \theta_0, M, \hat{\alpha})$ takes the explicit form:
\begin{equation}
\begin{split}
b_n =& \frac{2 r_{ph}^3(r_S - r_{ph})(r_{ph}^4 - 9 M^2 \hat{\alpha}) }{r_S\left(-2 M r_{ph}^3 + r_{ph}^4 + M^2 \hat{\alpha}\right)^{3/2}} \\
&\times \exp \Bigg[ \frac{\sqrt{r_{ph}^4 - 9 M^2 \hat{\alpha}}}{r_{ph}^2} \Bigg( k_O + k_S - \pi(n+1) \\
 &+ \arccos\left[ \frac{\sin \varphi}{\sqrt{\cot^2 \theta_0 + \sin^2 \varphi}} \right] \Bigg) \Bigg] \\
&+ \frac{r_{ph}^3}{\sqrt{-2 M r_{ph}^3 + r_{ph}^4 + M^2 \hat{\alpha}}},
\end{split}
\label{eq:bn_explicit}
\end{equation}
where the integration constants $k_O$ and $k_S$, associated with the observer and the source respectively, are evaluated with $\eta = 1 - r_{ph}/r$ as
\begin{equation}
   k_O = \int_{0}^{1 - \frac{r_{ph}}{r_0}} \mathcal{G}(\eta) d\eta, \qquad k_S = \int_{0}^{1 - \frac{r_{ph}}{r_S}} \mathcal{G}(\eta) d\eta,
   \label{eq:k_constants}
\end{equation}
and $\mathcal{G}=\mathcal{G}(\eta)$ given by
\begin{equation}
   \mathcal{G} = \frac{r_{ph}^2}{\sqrt{r_{ph}^4 - 9 M^2 \hat{\alpha}}} \sqrt{\frac{9 M^2 \hat{\alpha} - r_{ph}^4}{\eta \mathcal{Q}}} - \frac{r_{ph}^2}{|\eta| \sqrt{r_{ph}^4 - 9 M^2 \hat{\alpha}}},
   \label{eq:g_eta}
\end{equation}
\begin{equation}
\begin{split}
   \mathcal{Q} = & \, r_{ph}^4 (\eta-2) + 2 M r_{ph}^3 (\eta^2-3\eta+3) \\
   &+ M^2 \hat{\alpha} (\eta-2)(\eta^2-3\eta+3)(\eta^2-\eta+1).
\end{split}
   \label{eq:P_eta}
\end{equation}
Our result of Eq.~(\ref{eq:bn_explicit}) serves as an application example of Eq.~(26) from \cite{Aratore2024}. For the qOS spacetime, the integrals in Eq.~(\ref{eq:k_constants}) must be evaluated numerically. We have verified that in the Schwarzschild limit ($\hat{\alpha} \to 0$), the expression in Eq.~(\ref{eq:bn_explicit}) matches exactly the analytical forms given in Eq.~(36) of \cite{Aratore2024} and Eq.~(18) of \cite{Tsupko_2022}. For a source configuration satisfying $r_{ph} < r_S < r_0$, Eq.~(\ref{eq:bn_explicit}) holds under the condition that light rays decrease in radius to a minimum distance $R_t$ before traveling toward the observer. However, when the source at $r_S$ is located very close to $r_{ph}$, light rays up to a high index $n$ do not experience a radial turning point and instead propagate directly outward with a monotonically increasing radius. Therefore, to ensure the validity of this formulation, we position $r_S$ sufficiently far from the photon sphere so that an inner turning point $R_t$ exists across all calculated light rays. Additionally, by substituting $M = M(S, \hat{\alpha})$ from Eq.~\eqref{eq:Mass_Entropy}, we can express the impact parameter $b_n = b_n(n, \varphi, r_S, r_0, \theta_0, S, \hat{\alpha})$ in the entropy space, allowing us to identify the Davies point for each higher-order photon ring, as illustrated in Fig.~\ref{Shadows9} for $n=2$. 

Table \ref{tab:combined-model-I} compares the semi-analytical and numerical ray-tracing results for the impact parameters $b_n$ in the qOS spacetime, considering the Profile~1 thin accretion disk from Eq.~\eqref{profile1}, bounded as $r_S \in [r_{ISCO}, \infty)$. By assuming an accretion disk with specific inner and outer radii, the corresponding $n$-th photon ring will also have inner $b_n^{in}$ and outer $b_n^{out}$ boundaries determined by these emission limits. Consequently, the apparent thickness of the $n$-th photon ring is defined as $\delta b_n \equiv b_n^{out} - b_n^{in}$. We find excellent agreement between the semi-analytical and numerical ray-tracing results, with virtually identical values up to the fourth decimal place for $n\ge3$. Higher-order images are exponentially demagnified, with rapidly converging boundaries; for $n=4,5,6$, $\delta b_n$ is negligible, producing ultra-thin rings near the critical shadow radius: $r_{sh}\approx5.00214M$ for extremal qOS black holes and $r_{sh}\approx5.19615M$ for Schwarzschild. The numerical approach also captures the dominant lower-order emissions, including the direct image ($b_0$), for which $b_0^{out}\to\infty$ due to the unbounded model, and the broader $n=1$ ring with significantly larger $\delta b_1$.

In both Parameter and Entropy Spaces, the emission boundaries of these impact parameters, including their Davies point values $b_{nm}$, are constrained by black hole thermodynamics. In the Entropy Space, $b_n \to \infty$ as $S \to \infty$. This occurs because $b_n$ is expressed in units of $\sqrt{\hat{\alpha}}$. For fixed $\hat{\alpha}$, $S \to \infty$ implies $M \to \infty$ and $\hat{\alpha}/M^2 \to 0$, recovering the Schwarzschild limit. Hence, $b_n/\sqrt{\hat{\alpha}} \to \infty$. However, for any finite, non-zero $\hat{\alpha}$ valid within the black hole regime, the inner and outer boundaries $b_n^{in}$ and $b_n^{out}$ are perfectly well-defined. For an extremal qOS black hole in Entropy Space, the $n=6$ ring has negligible thickness, $\delta b_6\approx0$, with $b_6^{in}\approx b_6^{out}\approx r_{sh}\approx3.85065\sqrt{\hat{\alpha}}$. At the Davies point, these become $b_{6m}\approx5.05602M$ in Parameter Space and $4.46283\sqrt{\hat{\alpha}}$ in Entropy Space. This mapping of impact parameters into the Entropy Space introduces a novel thermodynamic perspective on future observational measurements of photon rings. For example, according to Table \ref{tab:combined-model-I}, if future high-resolution observations measure the wider $n=1$ lensed ring with an inner boundary in the range $b_1^{in} \in (5.34570M, 5.38018M)$ and an outer boundary $b_1^{out} \in (6.09538M, 6.11323M)$, or equivalently, $b_1^{in} \in (4.11512\sqrt{\hat{\alpha}}, 4.74896\sqrt{\hat{\alpha}})$ in the Entropy Space, these observational values would fall exactly between the extremal limit and the Davies point, which would suggest that the observed compact object is a thermodynamically stable qOS black hole.

\subsection{Gap Parameters}\label{sec3.4}

The angular size of a single photon ring, $\theta_n=b_n/r_0$, is degenerate with $M$, the observer distance, and the emission radius $r_S$. Thus, one ring cannot uniquely distinguish spherically symmetric spacetimes. To break this degeneracy, we introduce the gap parameter $\Delta_n$, defined as the relative separation between consecutive photon rings, normalized by the larger radius \cite{Aratore2024}:


\begin{figure*}[h]
 \vspace{0.2cm}
\begin{minipage}[t]{0.475\linewidth}
 \centering
 \hspace{-7.2cm} \includegraphics[width=0.94\linewidth]{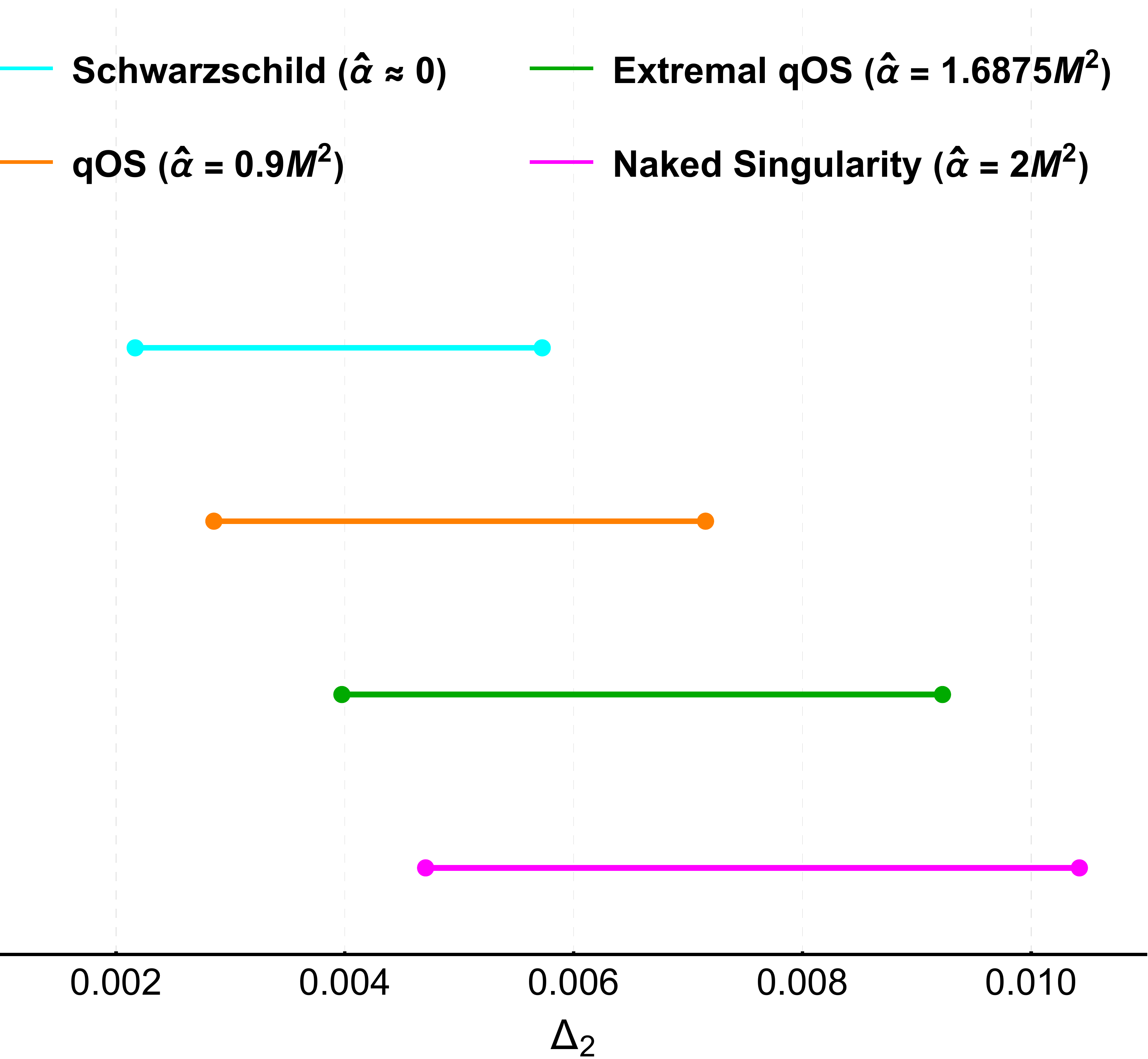}
  \hspace{-8cm} (a)
\end{minipage}%
\hfill%
\begin{minipage}[t]{0.49\linewidth}
 \centering
 \hspace{-7.7cm} \includegraphics[width=1.02\linewidth]{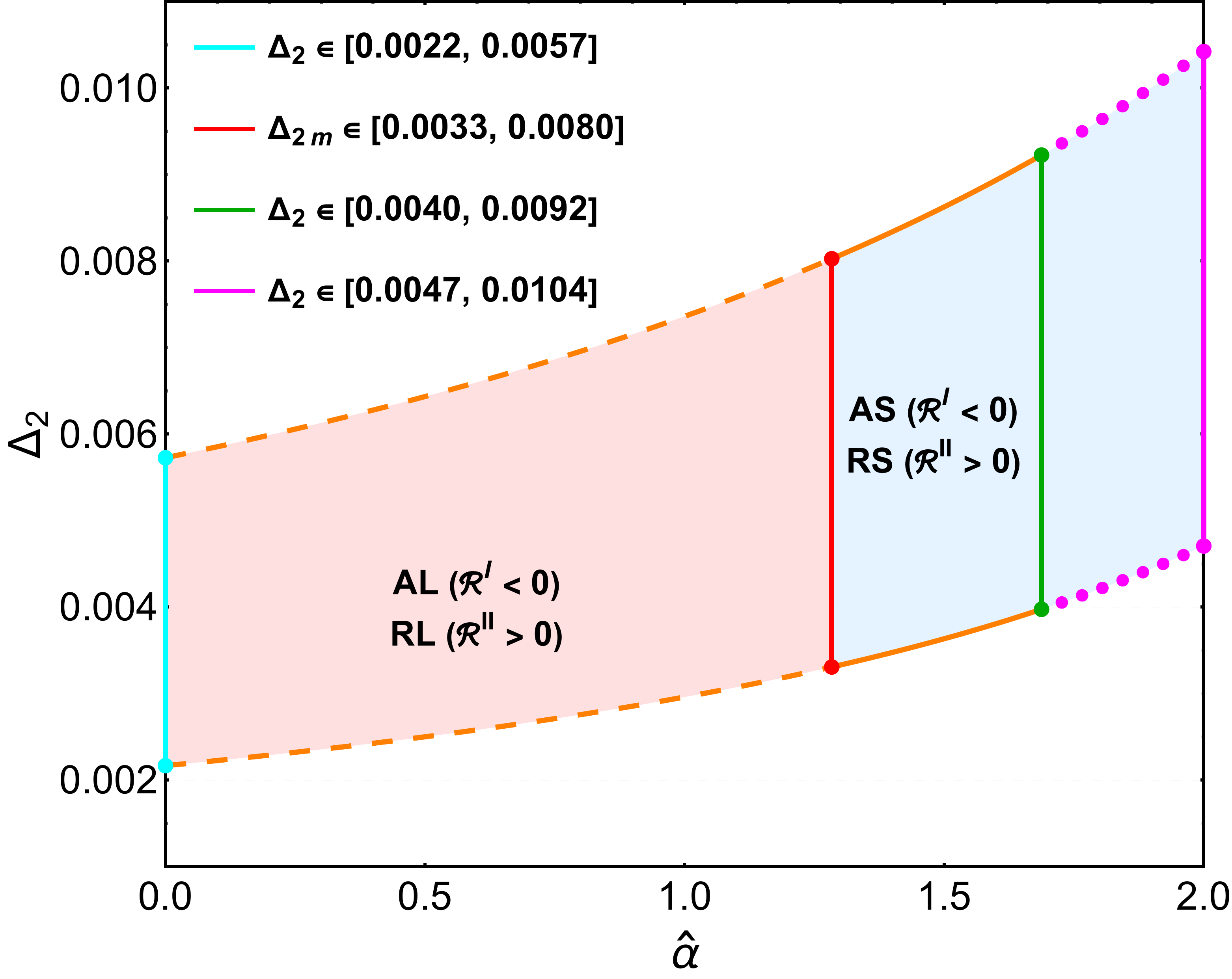}
 \hspace{-8cm} (b)
\end{minipage}%
\hfill \\
\vspace{-0.1cm}
\caption{Intervals of the gap parameter $\Delta_2$, measuring the relative separation between the $n=2$ and $n=3$ photon rings, obtained by varying $r_S$ from $r_{ISCO}$ to infinity using Eq. \eqref{eq:delta_n_explicit}. In (a), $\Delta_2$ is compared for Schwarzschild, qOS, and extremal qOS black holes, and naked singularities. In (b), $\Delta_2$ is shown versus $\hat{\alpha}$, with the corresponding MTPs; small phases are thermodynamically stable. AS, RS, AL, and RL denote Attractive Small, Repulsive Small, Attractive Large, and Repulsive Large phases, respectively.}

\label{Shadows10}
\end{figure*}


\begin{equation}
\Delta_n = \frac{\theta_n - \theta_{n+1}}{\theta_n} = \frac{b_n - b_{n+1}}{b_n}.
\label{deltan}
\end{equation}
Its main advantage is that, within the adopted approximations, it is independent of the observer's distance $r_0$. Using the explicit impact parameter $b_n$, we derive the analytical gap parameter $\Delta_n=\Delta_n(n,r_S,M,\hat{\alpha})$ for higher-order rings ($n\ge2$), evaluated along the horizontal axis $\varphi\in\{0^\circ,180^\circ\}$, where it is independent of $\theta_0$ (equivalently, for a polar observer, $\theta_0=0$):

\begin{equation}
\begin{split}
\Delta_n &= 2(r_S - r_{ph})(r_{ph}^4 - 9 M^2 \hat{\alpha}) \\
&\quad \times \Bigg\{ 2 (r_S - r_{ph})(r_{ph}^4 - 9 M^2 \hat{\alpha}) \\
&\quad\ \ \times \exp\!\left[ \frac{\sqrt{r_{ph}^4 - 9 M^2 \hat{\alpha}}}{r_{ph}^2} \big( k_O + k_S - \pi(n + \tfrac{1}{2}) \big) \right] \\
&\quad\ \ + r_S(-2 M r_{ph}^3 + r_{ph}^4 + M^2 \hat{\alpha}) \Bigg\}^{\!-1} \\
&\quad \times \Bigg\{ \exp\!\left[ \frac{\sqrt{r_{ph}^4 - 9 M^2 \hat{\alpha}}}{r_{ph}^2} \big( k_O + k_S - \pi(n + \tfrac{1}{2}) \big) \right] \\
&\quad\ \ - \exp\!\left[ \frac{\sqrt{r_{ph}^4 - 9 M^2 \hat{\alpha}}}{r_{ph}^2} \big( k_O + k_S - \pi(n + \tfrac{3}{2}) \big) \right] \Bigg\}.
\end{split}
\label{eq:delta_n_explicit}
\end{equation}

As shown in Table~\ref{tab:gap-parameters-unified}, the semi-analytical results from Eq.~\eqref{eq:delta_n_explicit} are invalid for the direct ($n=0$) and secondary ($n=1$) images, since the strong-deflection logarithmic approximation requires at least one complete photon orbit. Thus, $\Delta_0$ and $\Delta_1$ are obtained only by numerical ray tracing. For $n=2$, the semi-analytical method agrees closely with numerical results, with relative errors of 0.4--1.2\%, consistent with other spacetimes \cite{Aratore2024}.
As $n$ increases, $\Delta_n$ decreases exponentially, with higher-order rings approaching the critical impact parameter $r_{sh}$, so $\Delta_4,\Delta_5\approx0$ and $\Delta_n\to0$ as $n\to\infty$. For $n\ge3$, the error is 0.0\%, confirming the semi-analytical approximation for higher-order qOS images. Table~\ref{tab:gap-parameters-unified} links photon-ring observables to the thermodynamic structure, with the Davies point ($T=T_m$) separating stable and unstable branches where $C_{\hat{\alpha}}$ diverges. Thus, the Davies-point gap provides an observable signature of thermodynamic instability. Since $\Delta_n$ spans from $T=0$ to the classical limit $T=T_{\text{Schw}}$, future observations could use these separations to probe the thermodynamic phase. In particular, comparing $\Delta_1$ with $\Delta_{1m}$ identifies the stability branch, while $\Delta_{5m}\approx0.00000$ indicates that measurable thermodynamic information is mainly encoded in the first few photon-ring images.

Evaluating $\Delta_n$ over all emission radii $r_S\in[r_{ISCO},\infty)$ yields a bounded, intrinsic qOS gap interval \cite{Aratore2024}. Fig.~\ref{Shadows10}(a) shows this for $\Delta_2$ across different spacetime configurations, linking macroscopic features to GTD and microscopic interactions (Section~\ref{sec4}). As $\hat{\alpha}$ increases, the gap interval shifts monotonically to higher values.
For Schwarzschild ($\hat{\alpha}=0$), $\Delta_2\in[0.0022,0.0057]$, consistent with \cite{Aratore2024,Tsupko_2022}. For extremal qOS ($\hat{\alpha}=1.6875M^2$), $\Delta_2\in[0.0040,0.0092]$, while for a naked singularity ($\hat{\alpha}=2M^2$), $\Delta_2\in[0.0047,0.0104]$. qOS naked singularities retain photon and anti-photon spheres up to $\hat{\alpha}=2.84766M^2$. Since the semi-analytical approach applies whenever photon rings exist, it also covers these cases and yields the maximum gap-parameter values. Furthermore, a qOS naked singularity can in principle be distinguished from a Schwarzschild black hole under the adopted geometric and emission assumptions. For instance, at $\hat{\alpha} = 2.32 M^2$, the resulting interval $\Delta_2 \in [0.0058, 0.0118]$ is strictly distinct from the Schwarzschild case.

The physical implications of these bounds are direct \cite{Aratore2024}. First, they serve as a test of pure gravity: because the gap parameter isolates the background spacetime geometry from astrophysical variables such as the extent of the accretion disk, it provides a robust, mass-independent test of General Relativity. Second, they serve as an observational exclusion criterion: if future Very Long Baseline Interferometry missions measure the relative separation between two consecutive photon rings and the resulting gap parameter falls outside this theoretically predicted interval, the qOS metric can be conclusively ruled out by observation, even without precise knowledge of the emission radius.

\section{Geometrothermodynamics}\label{sec4}


Thermodynamic geometry describes the equilibrium space of a thermodynamic
system as a Riemannian manifold whose curvature encodes the effective
microscopic interactions
\cite{amari2012differential,weinhold1976metric,ruppeiner1979thermodynamics,
ruppeiner1995riemannian,ruppeiner2008thermodynamic}. GTD was introduced in \cite{quevedo2007geometrothermodynamics} to provide a description invariant under Legendre transformations, remaining independent of the chosen thermodynamic potential  \cite{quevedo2007geometrothermodynamics,
bravetti2014representation,azreg2014geometrothermodynamics}. In this framework, curvature singularities identify thermodynamic critical behaviors like Davies points, spinodal points, or phase transitions, offering deeper insights into the system's phase structure and instabilities. GTD has been successfully utilized in a wide array of physical systems \cite{quevedo2022geometrothermodynamics,quevedo2011phase,quevedo2023geometrothermodynamic,romero2026,zaldivar2023ideal,QuevedoQuevedoSanchez2026,BravettiLopezMonsalvoNettelQuevedo2014,quevedo2014geometric,2023IJGMM..2050057Q,e28070762,romero2026quasi,e25060944,PhysRevD.86.063508,2019EPJC...79..577B,2014GrCo...20..208Q,Sanchezz2026}, including, of course, black hole thermodynamics across numerous alternative theories of gravity \cite{quevedo2008geometrothermodynamics,Ladino:2024ned,ladino2025phase,LarranagaMojica2011,Gogoi,larranaga2012geometric,ref54,ref39,TharanathSureshKuriakose2015,Zhang2018GTD,ChannuieMomeni2018,GhaffariLuciano2025,Kurihara2018,Sanchez2016,QuevedoQuevedoSanchez2020,HuChenWang2017,GanChenWang2016,2025IJGMM..2250049J,SekhmaniEtAl2025,TajQuevedo2012,MoLiu2014,QuevedoSanchez2009,LucianoSheykhi2023,LucianoSaridakis2023,2023EPJC...83..710R,Beissen2024,JankeJohnstonKenna2010,quevedo2016geometrothermodynamics,HendiNaderi2015,HanChen2012,BravettiMomeniMyrzakulovAltaibayeva2013,Quevedo:2011fk,RizwanKumaraRajaniVaidAjith2019,quevedo2016einstein,ahmed2026shadow,Ladino:2026dru,quevedo2019quasi,Maya2026,Hernando2026,SHAH2026117662,Baghbani2026,QUEVEDO2026117321,Alessa2026,Biswas2026,Gogoi2026,BiswasSatyajit,QUEVEDO2026170601}.

We apply GTD to the qOS black hole using the thermodynamic results in
Sec.~\ref{sec:thermo}: the fundamental equation $\Phi=M$ given in Eq.~\eqref{eq:Mass_Entropy}, the
first law $dM=T\,dS+A\,d\hat\alpha$ with the conjugate pair
Eqs.~\eqref{eq:conjugates}, the quasi-homogeneity coefficients Eqs.~\eqref{eq:betas}, the
Smarr relation Eq.~\eqref{eq:smarr} and the response functions
Eqs.~\eqref{eq:Calpha}--\eqref{eq:CA}. The equilibrium space is therefore
two-dimensional, with $E^{a}=(S,\hat\alpha)$.

GTD implements Legendre invariance by introducing the auxiliary phase space
$\mathcal{T}$, a $(2i+1)$-dimensional manifold with coordinates
$Z^{A}=\{\Phi,E^{a},I_{a}\}$, $a=1,\dots,i$, where $i$ is the number of
thermodynamic degrees of freedom. The line elements $G=G_{AB}dZ^{A}dZ^{B}$ of
the metrics of $\mathcal{T}$ are
\cite{quevedo2007geometrothermodynamics, quevedo2011phase, quevedo2023unified}
\begin{equation}
	G^{I/II}=\left(d\Phi-I_{a}dE^{a}\right)^{2}
	+\left(\xi_{ab}E^{a}I^{b}\right)\left(\chi_{cd}\,dE^{c}dI^{d}\right),
\end{equation}
\begin{equation}
	G^{III}=\left(d\Phi-I_{a}dE^{a}\right)^{2}
	+\sum_{a=1}^{i}\xi_{a}\left(E_{a}I_{a}\right)^{2k+1}dE^{a}dI^{a},
\end{equation}
with $\delta^{c}_{a}={\rm diag}(1,\dots,1)$, $\eta^{c}_{a}={\rm diag}(-1,\dots,1)$,
$\chi_{cd}=\delta_{cd}$ for $G^{I}$ and $\chi_{cd}=\eta_{cd}$ for $G^{II}$.
Pulling these back to the equilibrium space $\mathcal{E}$ through
$\varphi:E^{a}\mapsto\left(\Phi,E^{a},\partial\Phi/\partial E^{a}\right)$, and
choosing $\xi_{ab}=\beta_{ab}={\rm diag}(\beta_{1},\dots,\beta_{i})$,
$\xi_{a}=\beta_{a}$ and $k=0$, one obtains the induced metrics
\begin{equation}
	g^{I}_{ab}=\beta_{\Phi}\Phi\,\delta^{c}_{a}\frac{\partial^{2}\Phi}{\partial E^{b}\partial E^{c}},
	\qquad
	g^{II}_{ab}=\beta_{\Phi}\Phi\,\eta^{c}_{a}\frac{\partial^{2}\Phi}{\partial E^{b}\partial E^{c}},
	\label{eq:g1g2}
\end{equation}
\begin{equation}
	g^{III}=\sum_{a=1}^{i}\beta_{a}\left(\delta_{ad}E^{d}\frac{\partial\Phi}{\partial E^{a}}\right)
	\delta^{ab}\frac{\partial^{2}\Phi}{\partial E^{b}\partial E^{c}}\,dE^{a}dE^{c},
	\label{eq:g3}
\end{equation}
where the $\beta_{a}$ are the quasi-homogeneity coefficients of $\Phi=\Phi(E^{a})$ and the conformal factor $\beta_{\Phi}\Phi$ is
generated by the quasi-homogeneous Euler identity
\cite{quevedo2017homogeneity, quevedo2019quasi, quevedo2023unified,
Quevedoquasi}.

In the GTD interpretation, the sign of the curvature scalar $\mathcal{R}$ of the equilibrium
manifold measures the effective thermodynamic interaction between the
microscopic constituents of the black hole \cite{
ruppeiner1979thermodynamics, Ladino:2024ned, ladino2025phase}: $\mathcal{R} > 0$ indicates repulsive interactions, $\mathcal{R} < 0$ denotes attractive ones, and $\mathcal{R} = 0$ corresponds to non-interacting systems, such as an ideal gas with flat geometry. Combining this information
with the local stability inferred from the sign of $C_{\hat\alpha}$ in
Eq.~\eqref{eq:branches}, each state of the system can be assigned an Microscopic Thermodynamic Phase (MTP), following the classification
introduced in our previous analyses of charged and rotating black holes
\cite{Ladino:2024ned, ladino2025phase, Ladino:2026dru}. Combining the small/large
character of the branch with the sign of the curvature scalar, the possible
MTPs are: the Attractive Small (AS), Repulsive Small (RS), Noninteractive
Small (NS), Attractive Large (AL), Repulsive Large (RL) and Noninteractive
Large (NL) ones, where by Eq.~\eqref{eq:branches} the small phases are the
thermodynamically stable ones and the large phases the unstable ones.

\subsection{GTD Metrics}

To eliminate nested radicals in Eq.~\eqref{eq:Mass_Entropy}, we introduce $w\equiv\sqrt{r_{h}^{2}-\hat\alpha}$, which yields
\begin{equation}
	\hat\alpha=r_{h}^{2}-w^{2},\quad S=\pi r_{h}^{2},\quad
	\sigma=\sqrt{S-\pi\hat\alpha}=\sqrt{\pi}\,w ,
	\label{eq:wchart}
\end{equation}
and 
$M=r_{h}^{2}/(r_{h}+w)$. As $(r_{h},w)$ is merely a coordinate change on $\mathcal{E}$, curvature scalars remain chart-independent and metric components transform via $J^{a}{}_{i}=\partial E^{a}/\partial y^{i}$ with $y^{i}=(r_{h},w)$. Then, defining the dimensionless order parameter $z \equiv w/r_h = \sqrt{1-\hat{\alpha}/r_h^2}$ yields:
\begin{equation}
	\begin{split}
		z &\equiv \frac{w}{r_h}
		= \sqrt{1-\frac{\hat{\alpha}}{r_h^2}},
		\qquad
		r_h = \frac{\sqrt{\hat{\alpha}}}{\sqrt{1-z^2}},
		\\
		\qquad & M = \frac{r_h}{1+z},
		\qquad
		S = \frac{\pi\hat{\alpha}}{1-z^2}.
	\end{split}
	\label{eq:orderparam}
\end{equation}
Via Eqs.~\eqref{eq:extremal} and \eqref{eq:davies}, $z=1/2$ marks the extremal case ($T=0$), $z\to1$ the Schwarzschild limit ($\hat\alpha\to0$), and $z_m=(1+\sqrt{13})/6 \simeq 0.767592$ the Davies point, defining the physical interval $z\in[1/2,1)$. Then, the conjugate variables~\eqref{eq:Temperature} become
\begin{align}
	T&=\frac{2w-r_{h}}{2\pi w\,(r_{h}+w)}=\frac{2z-1}{2\pi r_{h}z(1+z)},\\
	A&=\frac{r_{h}^{2}}{2w\,(r_{h}+w)^{2}}=\frac{1}{2r_{h}z(1+z)^{2}}.
	\label{eq:TA}
\end{align}
The second potential derivatives~\eqref{eq:MSS}--\eqref{eq:Maa} read
\begin{align}
	M_{,SS}&=\frac{r_{h}^{2}+r_{h}w-3w^{2}}{4\pi^{2}r_{h}w^{3}(r_{h}+w)}
	=\frac{1+z-3z^{2}}{4\pi^{2}r_{h}^{3}z^{3}(1+z)},\\
	M_{,S\hat\alpha}&=-\frac{r_{h}^{2}+2r_{h}w-2w^{2}}{4\pi w^{3}(r_{h}+w)^{2}}
	=-\frac{1+2z-2z^{2}}{4\pi r_{h}^{3}z^{3}(1+z)^{2}},\\
	M_{,\hat\alpha\hat\alpha}&=\frac{r_{h}^{2}(r_{h}+3w)}{4w^{3}(r_{h}+w)^{3}}
	=\frac{1+3z}{4r_{h}^{3}z^{3}(1+z)^{3}} .
	\label{eq:secondderiv}
\end{align}
The response functions Eqs.~\eqref{eq:Calpha}--\eqref{eq:CA} take the form
\begin{align}
	C_{\hat\alpha}&=\frac{2\pi r_{h}^{2}z^{2}(2z-1)}{1+z-3z^{2}},\label{eq:Calphaz}\\
	\kappa_{S}&=\frac{4r_{h}^{3}z^{3}(1+z)^{3}}{1+3z},\label{eq:kappaSz}\\
	C_{A}&=\frac{2\pi r_{h}^{2}(1-2z)(1+3z)}{z(1+4z)} ,\label{eq:CAz}
\end{align}
reproducing the extremality and Davies conditions of Sec.~\ref{sec:thermo}: the numerator of $C_{\hat\alpha}$ vanishes only at $z=1/2$ and its denominator at $3z^{2}-z-1=0$ ($z=z_{m}$), with $C_{\hat\alpha}\to-2\pi r_{h}^{2}=-2S$ in the Schwarzschild limit.

For $E^{1}=S$, $E^{2}=\hat\alpha$ and the weights~\eqref{eq:betas}, Eqs.~\eqref{eq:g1g2}--\eqref{eq:g3} yield the three line elements
\begin{align}
	g^{I}&=\beta_{M}M\left(M_{,SS}\,dS^{2}
	+2M_{,S\hat\alpha}\,dS\,d\hat\alpha
	+M_{,\hat\alpha\hat\alpha}\,d\hat\alpha^{2}\right),\label{eq:metric1}\\[2pt]
	g^{II}&=\beta_{M}M\left(-M_{,SS}\,dS^{2}
	+M_{,\hat\alpha\hat\alpha}\,d\hat\alpha^{2}\right),\label{eq:metric2}\\[2pt]
	g^{III}&=\Sigma_{S}M_{,SS}\,dS^{2}
	+\left(\Sigma_{S}+\Sigma_{\hat\alpha}\right)M_{,S\hat\alpha}\,dS\,d\hat\alpha
	\nonumber\\
	&\quad+\Sigma_{\hat\alpha}M_{,\hat\alpha\hat\alpha}\,d\hat\alpha^{2},\label{eq:metric3}
\end{align}
where $\Sigma_{S}=\beta_{S}ST=2ST$ and
$\Sigma_{\hat\alpha}=\beta_{\hat\alpha}\hat\alpha A=2A\hat\alpha$. In a 2D equilibrium space, the curvature singularity conditions \cite{quevedo2023unified, Ladino:2024ned} are:
\begin{align}
	I&:\quad \Phi_{,SS}\Phi_{,E^{2}E^{2}}-\left(\Phi_{,SE^{2}}\right)^{2}=0,
	\label{eq:condI}\\
	II&:\quad \Phi_{,SS}\Phi_{,E^{2}E^{2}}=0,\label{eq:condII}\\
	III&:\quad \Phi_{,SE^{2}}=0.\label{eq:condIII}
\end{align}
Evaluated for $\Phi=M$, $E^{2}=\hat\alpha$ via response functions~\eqref{eq:Cadef}--\eqref{eq:response}, these become
\begin{equation}
	I:\;\frac{T}{C_{A}\,\kappa_{S}}=0,
	\quad
	II:\;\frac{T}{C_{\hat\alpha}\,\kappa_{S}}=0,
	\quad
	III:\;\frac{1}{\alpha_{S}}=0 ,
	\label{eq:condresponse}
\end{equation}
enabling a direct test of GTD curvature singularities against response function divergences \cite{quevedo2011phase, quevedo2023unified}. Via Eq.~\eqref{eq:secondderiv}, the Hessian determinant Eq.~\eqref{eq:hessdet} is
\begin{equation}
	M_{,SS}M_{,\hat\alpha\hat\alpha}-\left(M_{,S\hat\alpha}\right)^{2}
	=-\frac{r_{h}+4w}{16\pi^{2}w^{3}(r_{h}+w)^{4}}<0 ,
	\label{eq:hessdetz}
\end{equation}
which remains strictly negative for all $r_{h},w>0$, showing condition $I$ is never satisfied in the mass representation. As $r_{h}+3w \neq 0$ in $M_{,\hat\alpha\hat\alpha}$, condition $II$ is satisfied uniquely at the Davies point:
\begin{equation}
	II:\;M_{,SS}=0 \iff 3z^{2}-z-1=0 \iff z=z_{m}.
\end{equation}
Condition $III$ ($M_{,S\hat\alpha}=0 \iff 2z^{2}-2z-1=0$) yields $z=(1+\sqrt3)/2\simeq1.366$, which lies outside the physical interval $z\in[1/2,1)$, rendering condition $III$ empty. Thus, the qOS black hole features a single spinodal point, unlike the richer phase structures of Kerr--AdS and RN--AdS solutions \cite{Ladino:2024ned, ladino2025phase}.

\subsection{The Enthalpy Representation}\label{sec:enthalpy}

Because black hole stability depends on the statistical ensemble \cite{caldarelli2000thermodynamics, quevedo2014ensemble}, the singularity conditions Eqs.~\eqref{eq:condI}--\eqref{eq:condIII} depend on the chosen thermodynamic potential. For a two-dimensional equilibrium manifold, $\mathcal{R}^{II}$ divergences track the chosen ensemble's response functions, while $\mathcal{R}^{I}$ diverges with the \emph{dual} representation's heat capacity \cite{Ladino:2026dru}. To correctly reproduce the singularities of the heat capacity $C_{E^{2}}$, the conditions are:
\begin{equation}
\begin{split}
I&:\; H_{,SS}H_{,I_{2}I_{2}}-\left(H_{,SI_{2}}\right)^{2}=0,\\
II&:\; M_{,SS}M_{,E^{2}E^{2}}=0 .
\end{split}
\label{eq:criterion}
\end{equation}
Thus, $g^{I}$ is constructed from the enthalpy $H(S,I_{2})$ and $g^{II}$ from the mass $M(S,E^{2})$ \cite{Ladino:2026dru}. We implement this criterion for the qOS black hole, setting $E^{2}=\hat\alpha$ and $I_{2}=A$. Via Legendre transformation $H\equiv M-A\hat\alpha$ and the Smarr relation Eq.~\eqref{eq:smarr}, the enthalpy is $H=2TS+A\hat\alpha$. With Eq.~\eqref{eq:wchart}, this takes the form
\begin{equation}
H(S,A)=\frac{r_{h}^{2}\left(3w-r_{h}\right)}{2w\left(r_{h}+w\right)}
=\frac{r_{h}\left(3z-1\right)}{2z(1+z)} .
\label{eq:enthalpy}
\end{equation}
The first law becomes $dH=T\,dS-\hat\alpha\,dA$, yielding the equations of state
\begin{equation}
T=\left(\frac{\partial H}{\partial S}\right)_{A},
\qquad
\hat\alpha=-\left(\frac{\partial H}{\partial A}\right)_{S} .
\label{eq:enthalpyEOS}
\end{equation}
The enthalpy remains quasi-homogeneous, with $\beta_{S}=2$, $\beta_{A}=\beta_{M}-\beta_{\hat\alpha}=-1$ and $ \beta_{H}=\beta_{M}=1$.
The associated Euler identity $\beta_{S}TS-\beta_{A}\hat\alpha A=\beta_{H}H$ recovers the Smarr formula Eq.~\eqref{eq:smarr}, confirming representational consistency. Notably, since $\beta_A<0$, the variable $A$ is not genuinely intensive \cite{Ladino:2026dru}. Moreover, the second derivatives of the enthalpy are
\begin{align}
	H_{,SS}&=-\frac{1+4z}{4\pi^{2}r_{h}^{3}(1+z)(1+3z)},\label{eq:HSS}\\
	H_{,SA}&=\frac{(1+z)\left(2z^{2}-2z-1\right)}{\pi(1+3z)},\label{eq:HSA}\\
	H_{,AA}&=-\frac{4r_{h}^{3}z^{3}(1+z)^{3}}{1+3z}=-\kappa_{S} .\label{eq:HAA}
\end{align}
From
Eq.~\eqref{eq:criterion} we obtain
\begin{equation}
	I:\; H_{,SS}H_{,AA}-\left(H_{,SA}\right)^{2}
	=\frac{(1+z)^{2}\left(3z^{2}-z-1\right)}{\pi^{2}(1+3z)} ,
	\label{eq:hessdetH}
\end{equation}
which vanishes solely at $z=z_{m}$ on the physical branch. Then, in the enthalpy representation, condition $I$ reproduces the divergence of $C_{\hat\alpha}$, agreeing with Eq.~\eqref{eq:criterion}. Consequently, the line element is
\begin{equation}
g^{I}=\beta_{H}H\left(H_{,SS}dS^{2}+2H_{,SA}dSdA
+H_{,AA}dA^{2}\right),
\label{eq:metric1H}
\end{equation}
giving the GTD metric for the $H(S,A)$ representation.

\begin{table*}[h]
	\centering
	\caption{GTD scalars for the qOS black hole on the physical branch $z\in[\tfrac12,1)$, and the resulting MTPs. }
	\label{tab:GTDsummary}
	\setlength{\tabcolsep}{6pt}
            \renewcommand{\arraystretch}{1.1} 
	\begin{tabular*}{\textwidth}{@{\extracolsep{\fill}}l|l|l|l|l|l}
		\hline
		\textbf{Metric} & \textbf{Potential} & \textbf{Scalar} & \textbf{Singularities} & \textbf{Microstructure} & \textbf{MTPs} \\
		\hline
		$g^{I}$   & $H(S,A)$ & $\mathcal{R}^{I}<0$ & $S=S_{m}$ & Attractive & AS $\to$ AL \\
		$g^{II}$  & $M(S,\hat\alpha)$ & $\mathcal{R}^{II}>0$ & $S=S_{m}$ & Repulsive & RS $\to$ RL \\
		\hline
		$g^{I}$ (not selected)  & $M(S,\hat\alpha)$ & $\mathcal{R}^{I}_{M}\equiv0$ (flat) & None & No interaction & NS/NL  \\
		$g^{III}$ (not selected) & $M(S,\hat\alpha)$ & $\mathcal{R}^{III}<0$ & None & Attractive & AS/AL  \\
		\hline
	\end{tabular*}
\end{table*}

\subsection{GTD Scalars}

Computing the Ricci scalars via Eqs.~\eqref{eq:orderparam}, first, for metric $g^{I}$ of Eq.~\eqref{eq:metric1H} we find
\begin{equation}
	\begin{split}
	\mathcal{R}^{I}&=-\frac{3z^{3}(1+z)^{3}\,q(z)}
	{\beta_{H}\,r_{h}^{2}\,(3z-1)^{3}(1+3z)^{2}\left(3z^{2}-z-1\right)^{2}},\\
	q(z)&=324z^{6}+81z^{5}-216z^{4}-3z^{3}+24z^{2}+8z+2 .
	\end{split}
	\label{eq:RI}
\end{equation}
Its unique physical pole is $3z^{2}-z-1=0$ ($z=1/3$ lies below extremality), so
\begin{equation}
	\mathcal{R}^{I}\to\infty
	\iff
	C_{\hat\alpha}\to\infty
	\iff
	z=z_{m} .
\end{equation}
Since $q(z)$ has no real roots in $z\in[\tfrac12,1)$ and $(3z-1)^{3}>0$,
\begin{equation}
	\mathcal{R}^{I}<0
	\qquad\text{throughout}\qquad
	z\in\left[\tfrac{1}{2},1\right) ,
\end{equation}
with extremal value $\hat\alpha\,\mathcal{R}^{I}=-4941/400=-12.3525$. Thus, $g^{I}$ assigns an attractive microstructure to the qOS black hole on both branches. For the metric $g^{II}$ of Eq.~\eqref{eq:metric2}, we find
\begin{equation}
	\begin{split}
	\hat\alpha\,\mathcal{R}^{II}
	&=\frac{\left(1-z^{2}\right)z^{2}(1+z)^{3}\,p(z)}
	{\beta_{M}\,(1+3z)^{2}\left(3z^{2}-z-1\right)^{2}},\\
	p(z)&=54z^{4}+45z^{3}-18z^{2}-4z+2.
	\end{split}
	\label{eq:RIIscaled}
\end{equation}
Since $(1+3z)^{2}$ has no positive roots, $\mathcal{R}^{II}\to\infty$ if and only if $C_{\hat\alpha}\to\infty$ at $z=z_{m}$, avoiding any spurious singularities. This guarantees both metrics diverge at the same physical state, fulfilling Eq.~\eqref{eq:criterion}. Furthermore, $p(z)$ has no physical roots and all remaining factors are positive. Consequently, 
\begin{equation}
		\mathcal{R}^{II}>0
		\qquad\text{throughout}\qquad
		z\in\left[\tfrac{1}{2},1\right) ,
	\end{equation}
maintaining its sign across the Davies point, with an extremal value of $\hat\alpha\,\mathcal{R}^{II}=81/100$. Therefore, the qOS black hole microstructure is always \emph{repulsive} on both branches, with a divergence at $z=z_{m}$.
As in the Reissner--Nordström and Kerr solutions \cite{Ladino:2026dru}, both metrics agree on the instability's location but disagree on the interaction type ($\mathcal{R}^{I}$ indicates attractive phases, $\mathcal{R}^{II}$ repulsive). Unlike Kerr, the qOS black hole lacks a noninteractive state ($\mathcal{R}^{II}=0$). Crucially, constructing $g^{I}$ from mass instead of enthalpy yields a trivially flat metric ($\mathcal{R}^{I}_{M}\equiv 0$), even though $\det g^{I}_{M} \neq 0$. This flatness is a direct consequence of the two-dimensional scaling structure ($\beta_1=\beta_2$) in Eq.~\eqref{eq:betas}, analogous to the Ruppeiner flatness of the Reissner--Nordström solution \cite{mirza2007ruppeiner, ruppeiner2008thermodynamic}. Therefore, the enthalpy representation dictated by Eq.~\eqref{eq:criterion} is essential to make $g^{I}$ informative. Finally, for the third metric we obtain
\begin{equation}
	\begin{split}
	\mathcal{R}^{III}
	&=-\frac{24\,z^{5}(1+z)^{3}\,N(z)}{r_{h}^{2}\,\Pi(z)^{2}},\\
	N(z)&=100z^{5}-120z^{4}-48z^{3}+88z^{2}-9z-6,\\ 
	\Pi(z)&=4z^{6}-80z^{5}+108z^{4}-39z^{2}+4z+4.
	\end{split}
	\label{eq:RIII}
\end{equation}
Because neither $\Pi$ nor $N$ has roots within the physical range $z\in[\tfrac12,1)$, we find that
\begin{equation}
	\mathcal{R}^{III}<0
	\quad\text{and regular for all}\quad
	z\in\left[\tfrac{1}{2},1\right) ,
\end{equation}
with the extremal value $\hat\alpha\,\mathcal{R}^{III}=-27/4$. Although well behaved, $\mathcal{R}^{III}$ fails to detect the Davies point and is thus discarded, consistent with Eq.~\eqref{eq:criterion}. Table~\ref{tab:GTDsummary} summarizes the results for the GTD scalars and the corresponding MTPs of each branch of the qOS black holes. It is worth highlighting that, following Eq.~\eqref{eq:criterion}, the selected metrics $g^{I}$ (from enthalpy $H(S,A)$) and $g^{II}$ (from mass $M(S,\hat\alpha)$) diverge at the Davies point $S_{m}$, where the stable small phase ($S<S_{m}$) shifts to the large phase ($S>S_{m}$).

\begin{table*}[h]
	\centering
	\caption{qOS black hole quantities on the physical branch $z\in[\tfrac12,1)$, for $\beta_{H}=\beta_{M}=1$.  The
		last column lists the MTPs predicted by
		$\mathcal{R}^{I}$ and $\mathcal{R}^{II}$, respectively.}
	\label{tab:GTDnumbers}
	\scriptsize
	\setlength{\tabcolsep}{2pt}
    \renewcommand{\arraystretch}{1.1} 
	\begin{tabular*}{\textwidth}{@{\extracolsep{\fill}}l|c|c|c|c|c|c|c|c|c|c|c}
		\hline
		$z$ & $r_{h}/\sqrt{\hat\alpha}$ & $M/\sqrt{\hat\alpha}$
		& $S/\hat\alpha$ & $T\sqrt{\hat\alpha}$
		& $C_{\hat\alpha}/\hat\alpha$ & $C_{A}/\hat\alpha$
		& $\kappa_{S}/\hat\alpha^{3/2}$
		& $\hat\alpha\,\mathcal{R}^{I}$ & $\hat\alpha\,\mathcal{R}^{II}$
		& $r_{sh}/\sqrt{\hat\alpha}$ & MTPs \\
		\hline
		$1/2$ (ext.) & $1.1547$ & $0.7698$ & $4.1888$ & $0$ & $0$ & $0$ & $1.0392$ & $-12.3525$ & $0.8100$ & $3.8506$ & AS / RS \\
		$0.55$ & $1.1974$ & $0.7725$ & $4.5041$ & $0.0156$ & $0.4241$ & $-1.3563$ & $1.6054$ & $-10.4882$ & $1.8385$ & $3.8654$ & AS / RS \\
		$0.60$ & $1.2500$ & $0.7813$ & $4.9087$ & $0.0265$ & $1.3593$ & $-2.6950$ & $2.4686$ & $-12.8544$ & $4.3797$ & $3.9133$ & AS / RS \\
		$0.70$ & $1.4003$ & $0.8237$ & $6.1600$ & $0.0382$ & $10.499$ & $-5.7431$ & $5.9701$ & $-69.3551$ & $45.356$ & $4.1444$ & AS / RS \\
		$z_{m}=0.7676$ & $1.5602$ & $0.8827$ & $7.6474$& $0.0402$ & $\pm\infty$ & $-8.6530$ & $11.4886$ & $-\infty$ & $+\infty$ & $4.4628$ & AS\,$\to$\,AL / RS\,$\to$\,RL \\
		$0.80$ & $1.6667$ & $0.9259$ & $8.7267$ & $0.0398$ & $-55.851$ & $-10.597$ & $16.2635$ & $-319.305$ & $261.84$ & $4.6949$ & AL / RL \\
		$0.90$ & $2.2942$ & $1.2075$ & $16.5347$ & $0.0325$ & $-40.432$ & $-23.644$ & $65.2704$ & $-15.8024$ & $14.289$ & $6.1893$ & AL / RL \\
		$0.95$ & $3.2026$ & $1.6423$ & $32.2215$ & $0.0241$ & $-69.101$ & $-48.968$ & $216.954$ & $-5.3039$ & $4.9495$ & $8.4733$ & AL / RL \\
		$0.99$ & $7.0888$ & $3.5622$ & $157.869$ & $0.0112$ & $-319.13$ & $-250.16$ & $2744.45$ & $-0.8605$ & $0.8200$ & $18.4826$ & AL / RL \\
		\hline
	\end{tabular*}
\end{table*}

The comparison of the three GTD metrics, therefore, leads to a clear 
geometric characterization of the thermodynamic critical behavior. 
Within the physical domain, the relevant singularity of $g^{II}$ 
coincides with the divergence of the heat capacity at the Davies 
point, whereas the singularity condition associated with $g^{III}$ does not provide additional information 
on this thermodynamic transition. The corresponding analysis using $g^{I}$ 
in the enthalpy representation also reproduces the thermodynamic 
singularity. Thus, the Davies transition is consistently identified 
by the GTD metrics associated with the relevant thermodynamic 
stability conditions, independently of the subsequent interpretation 
of the signs of their curvature scalars.

\subsection{Microstructure and Shadows}

In the GTD analysis considered here, microscopic interaction 
properties are understood in an effective thermodynamic sense. 
No specific microscopic constituents or statistical model are 
assumed for the qOS black hole. Following the thermodynamic-curvature 
interpretation of Ref.~\cite{ruppeiner2010thermodynamic}, the curvature sign is 
used as an indicator of the effective character of thermodynamic 
interactions, while its singularities characterize the critical 
structure of the equilibrium system. This interpretation should be 
understood separately for each GTD metric, since different metrics 
can assign different curvature signs to the same equilibrium state. 
Therefore, the optical quantities discussed below are related to 
these effective thermodynamic interaction sectors through their 
common dependence on the equilibrium state, rather than providing 
a direct probe of microscopic black-hole degrees of freedom.

Because the whole phase structure collapses to the single value $z=z_{m}$ of a scale-invariant parameter, every geometric observable of Sec.~\ref{sec3} takes a definite value at this point. Using Eqs.~\eqref{eq:photon_sphere} and \eqref{eq:shadow_radius}, the shadow radius $r_{sh}$ exhibits monotonic growth across the physical branch $z \in [1/2, 1)$. The common pole of $\mathcal{R}^{I}$ and $\mathcal{R}^{II}$ at the Davies point maps to
\begin{equation}
	\begin{split}
	r_{phm}&\simeq2.498095\sqrt{\hat\alpha}\simeq2.830138\,M ,\\
	r_{shm}&\simeq4.462827\sqrt{\hat\alpha}\simeq5.056019\,M ,
	\end{split}
	\label{eq:shadowdavies}
\end{equation}
to be compared with the extremal values
\begin{equation}
	\begin{split}
	r_{ph}^{\rm ext}&\simeq2.123825\sqrt{\hat\alpha}\simeq2.758930\,M ,\\
	r_{sh}^{\rm ext}&\simeq3.850649\sqrt{\hat\alpha}\simeq5.002140\,M ,
	\end{split}
	\label{eq:shadowextremal}
\end{equation}
which bound $r_{sh}$ from below, while the upper bound is given by the Schwarzschild limit $r_{sh}=3\sqrt3\,M\simeq5.196152\,M$. The values in Eq.~\eqref{eq:shadowdavies} and Eq.~\eqref{eq:shadowextremal} coincide with the entries $r_{phm}$, $r_{shm}$, the extremal endpoints of Table~\ref{tab:unified-observables}, and the $n\to\infty$ limit of the photon-ring impact parameters $b_{nm}$ of Table~\ref{tab:combined-model-I}. Thus, the GTD prediction for the shadow at the critical point agrees exactly with the divergence of the thermodynamic response function $C_{\hat\alpha}$ of Sec.~\ref{sec:thermo}. Overall, the shadow radius is constrained within $5.0021\,M \lesssim r_{sh} \lesssim 5.1962\,M$ in parameter space, which translates to $r_{sh}/\sqrt{\hat{\alpha}} \gtrsim 3.8506$ within the entropy space, where $r_{sh}/\sqrt{\hat{\alpha}}$ grows without bound as $S\to\infty$. To illustrate the evolution of these properties in detail, Table \ref{tab:GTDnumbers} compiles the numerical values of the main geometric and thermodynamic quantities of qOS black holes along the physical branch $z \in [1/2, 1)$. Since $\mathcal{R}^{I}<0$ and $\mathcal{R}^{II}>0$ throughout $z\in[1/2,1)$, a measurement of $r_{sh}$ places the qOS black hole into one of the following MTPs described by the shift at the Davies point:

\begin{itemize}
\item \textbf{Stable branch:} Below the Davies threshold, $5.0021\,M \le r_{sh}<5.0560\}M$ ($3.8506 \le r_{sh}/\sqrt{\hat{\alpha}}<4.4628$), corresponding to AS in $\mathcal{R}^{I}$ and RS in $\mathcal{R}^{II}$. For $C_{\hat\alpha}>0$, the branch is bounded by the extremal configuration Eq.~\eqref{eq:shadowextremal}, where curvature scalars remain finite and $T=C_{\hat\alpha}=C_A=0$, with finite microscopic interaction at zero temperature.

\item \textbf{The Davies point ($r_{sh} = 5.0560\,M$):} The threshold where $\mathcal{R}^{I}$, $\mathcal{R}^{II}$, and $C_{\hat\alpha}$ diverge simultaneously while $\kappa_{S}$ and $C_{A}$ stay finite. The microscopic interaction is strongest here, consistent with the loss of local thermal stability.

\item \textbf{Unstable branch:} Beyond the Davies point, $C_{\hat\alpha}<0$ and $5.0560\,M<r_{sh}\le5.1962\,M$ ($4.4628<r_{sh}/\sqrt{\hat{\alpha}}\mathrel{<}\infty$), corresponding to AL in $\mathcal{R}^{I}$ and RL in $\mathcal{R}^{II}$. The branch approaches the Schwarzschild limit, where curvature scalars remain finite for finite $r_h$ and vanish only as $r_h\to\infty$.

\end{itemize}

The exact critical shadow boundary of a black hole or naked singularity cannot be directly measured. However, as shown in Sec.~\ref{sec3}, impact parameters and photon-ring angular sizes can be extracted from shadow images. Table~\ref{tab:combined-model-I} shows the qOS impact parameters $b_n$ for a light source spanning $r_S\in[r_{ISCO},\infty)$ and their direct connection to the MTPs. For instance, if future high-resolution observations measure the $n=1$ photon ring with an inner boundary in the range $b_1^{in} \in (5.34570M, 5.38018M)$ and an outer boundary $b_1^{out} \in (6.09538M, 6.11323M)$, or equivalently in the Entropy Space, $b_1^{in} \in (4.11512\sqrt{\hat{\alpha}}, 4.74896\sqrt{\hat{\alpha}})$ and $b_1^{out} \in (4.69223\sqrt{\hat{\alpha}}, 5.39600\sqrt{\hat{\alpha}})$, these observational values would suggest that the observed compact object is a thermodynamically stable AS (for $\mathcal{R}^{I}<0$) or RS (for $\mathcal{R}^{II}>0$) qOS black hole.
On the contrary, if the observed impact parameters fall beyond the Davies point, yielding an inner boundary in the range $b_1^{in} \in (5.38018M, 5.47892M)$ and an outer boundary $b_1^{out} \in (6.11323M, 6.16755M)$, or equivalently $b_1^{in} \in (4.74896\sqrt{\hat{\alpha}}, \infty)$ and $b_1^{out} \in (5.39600\sqrt{\hat{\alpha}}, \infty)$ in the Entropy Space, this would indicate that the object can be a thermodynamically unstable AL (for $\mathcal{R}^{I}<0$) or RL (for $\mathcal{R}^{II}>0$) qOS black hole.

The MTPs of qOS can also be probed through photon-ring gap parameters $\Delta_n$. As shown in Fig.~\ref{Shadows10}(b), the $\Delta_2$ interval varies continuously with $\hat{\alpha}$ and shifts monotonically to higher values, showing that stronger quantum corrections increase the relative separation of higher-order photon rings. Relative to the Davies critical line, $\Delta_{2m}\in[0.0033,0.0080]$ defines distinct observational regimes. For $\Delta_2<0.0033$, the state lies unambiguously in the AL or RL phases, while $0.0080<\Delta_2\le0.0092$ identifies the stable AS or RS phases; values $\Delta_2>0.0092$ lie beyond the reach of any qOS black hole. Within the Davies interval, $\Delta_2\in[0.0033,0.0080]$, the state may be stable, unstable, or critical, depending on $r_S$ and the sign of the corresponding GTD curvature scalar. Naked singularities yield larger $\Delta_n$ values, but without an event horizon, the standard thermodynamic framework based on horizon area and surface gravity does not directly apply. Their optical properties remain well defined, but a consistent thermodynamic description likely requires a quantum-gravity framework in which the singularity is of Planck-scale size and carries only a few bits of entropy \cite{Xiang:2005wg}. Thus, their thermodynamics differs fundamentally from that of black holes and requires alternative entropy models, left for future work.

Larger shadows therefore correspond to large phases and smaller shadows to small phases, as in Reissner--Nordström and Kerr black holes \cite{Ladino:2026dru}. Consequently, a shadow measurement probes not only macroscopic parameters like $\hat\alpha$, but also the thermodynamical stability and the microscopic interaction type.

Two features are specific to qOS. First, for finite $r_h$, curvature scalars remain finite as $\hat\alpha\to0$, approaching $\mathcal R^I\to-165/(4\beta_H r_h^2)$, $\mathcal R^{II}\to79/(2\beta_M r_h^2)$, and $\mathcal R^{III}\to-960/r_h^2$, and vanish only as $r_h\to\infty$. GTD cannot uniquely recover this limit from Schwarzschild alone, whose equilibrium space lacks the structure to identify thermodynamic singularities \cite{mirza2007ruppeiner,Ladino:2026dru}. Unlike Schwarzschild (which lacks a Davies point), $\hat\alpha$ drives the spinodal structure, consistent with quantum-corrected cosmological spacetimes where quantum effects generate phase transitions \cite{romero2026quasi}. Second, unlike Reissner--Nordström--AdS
and Kerr--AdS cases \cite{Ladino:2024ned, ladino2025phase}, where $\mathcal{R}^{II}$ shifts from a repulsive small branch to an attractive large branch, here each scalar retains a constant sign across the transition. Thus, qOS black holes never alter their microscopic interaction character and admit no noninteractive state.

\section{Conclusions}\label{sec5}

    We investigated the optical and thermodynamic properties of qOS spacetimes across three regimes determined by $\hat{\alpha}/M^2$: black holes ($0<\hat{\alpha}/M^2<27/16$), extremal black holes ($\hat{\alpha}/M^2=27/16$), and naked singularities ($\hat{\alpha}/M^2>27/16$). We summarized their thermodynamics, including the phase structure, and response functions. We then analyzed test-particle motion and simulated shadows via numerical backward ray tracing with three thin accretion disk models to reveal their distinct optical features.
    Our analysis reveals a clear distinction between qOS black holes and qOS naked singularities. Unlike the dark shadows of qOS black holes, qOS naked singularities can exhibit central bright rings. This occurs because an infinite repulsive core barrier prevents photon trapping, rebounding light back to the observer. Despite measurable differences, qOS naked singularities effectively mimic black holes by reproducing key optical features like photon rings. Using numerical simulations for the qOS black hole cases, we calculated the specific possible intervals for the impact parameters $b_n$ for both the direct image and the photon rings. As $\hat{\alpha}$ increases, we found that the entire shadow size becomes smaller, encompassing the direct image, the photon rings, and consequently, the critical shadow boundary.   

Furthermore, we employed a semi-analytical strong-deflection approach to calculate higher-order photon-ring impact parameters $b_n$ ($n\ge2$) and gap parameters $\Delta_n$, complementing the numerical ray-tracing results. The two methods agree closely, with relative errors of 0.4--1.2\%. We find that $\Delta_n$ decreases exponentially with $n$, as higher-order rings asymptotically approach the shadow boundary. Evaluating $\Delta_n$ from $r_{ISCO}$ to infinity yields a bounded interval intrinsic to the qOS spacetime, relevant for future high-resolution observations. As $\hat{\alpha}$ increases, the gap interval shifts monotonically to higher values. Since qOS naked singularities can possess photon and anti-photon spheres up to $\hat{\alpha}=2.84766M^2$, the semi-analytical approach remains valid wherever photon rings exist, yielding the largest possible gap parameters. Consequently, a qOS naked singularity can be distinguished from a Schwarzschild black hole, as its $\Delta_n$ interval can be strictly distinct from the Schwarzschild. These theoretical bounds not only serve as a test of pure gravity but also provide an observational exclusion criterion.   

On the thermodynamic side, we applied GTD to qOS black holes and calculated their Ricci scalars, which capture the specific heat divergences at the Davies points. The metrics $g^I$ in the enthalpy representation and $g^{II}$ in the mass representation $M(S,\hat{\alpha})$ both reproduce these divergences. Our results reveal a connection between macroscopic optical observables and microscopic thermodynamic structure, with thermodynamic stability and attractive or repulsive MTP interactions encoded in the photon-ring signatures. By examining optical observables such as the gap parameters alongside the GTD interpretations of scalar behaviors, we classified these microscopic interactions based on the parameter $\hat{\alpha}$.   

The distinct optical features identified in this work, particularly the central bright rings, the overall smaller shadow sizes, and the higher possible gap parameter values of naked singularities, provide a theoretical perspective for distinguishing these exotic horizonless compact objects from black holes in future observations. Future work will explore the thermodynamics of naked singularities by extending GTD to their shadow and photon-ring signatures. We also propose applying this framework to compact objects in alternative gravity theories, such as nonconservative unimodular gravity~\cite{idrissov2026unimodular}, and black hole mimickers such as gravastars or wormholes \cite{Fauzi_2025,Rosa_2024,Chernov2023,Benavides2023}, whose geometries are nearly indistinguishable from those of standard black holes, making precise theoretical signatures essential to guide observational efforts.

\section*{Acknowledgments}
CRF and JML acknowledge support from SECIHTI-Mexico, grants No. 4003366 and No. 402076. AI acknowledges financial support from SECIHTI through the National Postgraduate Scholarship Program (CVU 2222058). This work was supported by UNAM-DGAPA-PAPIIT, grant No. 108225 and Conahcyt, grant No. CBF-2025-I-243.
\renewcommand{\addcontentsline}[3]{\oldaddcontentsline{#1}{#2}{#3}}
\appendix

\let\oldaddcontentsline\addcontentsline
\renewcommand{\addcontentsline}[3]{}
 \setlength{\bibsep}{0pt}
\bibliographystyle{unsrt}
\bibliography{sn-bibliography}
\end{document}